\documentclass[aip, amsmath, amssymb, reprint]{revtex4-2}

\usepackage{graphicx}
\usepackage{blindtext}
\usepackage{color}
\usepackage{dcolumn}
\usepackage{xcolor}
\usepackage{siunitx}

\begin{document}

% Setup for SIUnitX
\sisetup{per-mode = symbol}

% Title of paper
\title{Lattice Matching Dictates the Growth Mode and Quality of Deuterium Crystallization in Confined Spherical Shells}

% Authors
\author{Peng Bi}
\thanks{Author to whom correspondence should be addressed: bipeng010@swust.edu.cn, yiyong@swust.edu.cn, chenqf01@gmail.com.}
\affiliation{School of Mathematics and Physics, Southwest University of Science and Technology, Mianyang 621010, China}

\author{Yu-Shen Wan}
\thanks{Yu-Shen Wan and Peng Bi contributed equally.} 
\affiliation{Beijing National Laboratory for Condensed Matter Physics, Institute of Physics, Chinese Academy of Sciences, Beijing 100190, China}
\affiliation{College of Materials Science and Opto-Electronic Technology, University of Chinese Academy of Sciences, Beijing 100049, China}

\author{Wei Zhang}
\affiliation{School of Mathematics and Physics, Southwest University of Science and Technology, Mianyang 621010, China}

\author{Jian Chen}
\affiliation{School of Mathematics and Physics, Southwest University of Science and Technology, Mianyang 621010, China}

\author{Yong Yi}
\thanks{Author to whom correspondence should be addressed: bipeng010@swust.edu.cn, yiyong@swust.edu.cn, chenqf01@gmail.com.}
\affiliation{School of Materials and Chemistry, Southwest University of Science and Technology, Mianyang 621010, China}

\author{Qi-Feng Chen}
\thanks{Author to whom correspondence should be addressed: bipeng010@swust.edu.cn, yiyong@swust.edu.cn, chenqf01@gmail.com.}
\affiliation{School of Mathematics and Physics, Southwest University of Science and Technology, Mianyang 621010, China}
\affiliation{National Key Laboratory for Shock Wave and Detonation Physics Research, Institute of Fluid Physics, China Academy of Engineering Physics, Mianyang 621900, China}

% Date
\date{\today}

% Abstract
\begin{abstract}
Cryogenic hydrogen isotope fuel layers with high structural integrity and atomic-scale smoothness are prerequisites for symmetric implosion and ignition in inertial confinement fusion (ICF). Using deuterium (D$_2$) as model fuel, we perform large-scale molecular dynamics simulations with a Feynman-Hibbs corrected Silvera-Goldman potential to describe nuclear quantum effects at low temperatures, systematically investigating D$_2$ crystallization inside spherical ablator capsules. By varying substrate lattice constant from 3.1 Å to 3.9 Å, we demonstrate that lattice matching dictates the transition from coherent epitaxial growth to polycrystalline formation, establishing it as the primary design principle for high-performance targets. When the substrate lattice closely matches the equilibrium hexagonal-close-packed (HCP) spacing of cryogenic D$_2$ ($\sim$3.5 Å), D$_2$ forms coherent layer-by-layer epitaxial growth consistent with Ostwald’s stepwise nucleation theory, yielding HCP-dominated near-single crystals with minimal dislocations and ultra-smooth inner surfaces. In contrast, large lattice mismatch destabilizes coherent growth and causes island-like growth, producing polycrystalline structures with mixed HCP/FCC phases, elevated defects, and greatly increased surface roughness. Radial stress analysis shows that interfacial stress from mismatch localizes within 2–3 molecular layers near the interface, triggering subsequent defect-mediated growth. These findings highlight substrate lattice matching in regulating confined solid growth and crystallization quality, establish it as a key principle for ablator inner-surface engineering in ICF cryogenic targets, and offer atomic guidance for growing high-quality single-crystal deuterium-tritium (DT) fuel layers with optimal smoothness.
\end{abstract}

\maketitle

\section{Introduction}
Inertial Confinement Fusion (ICF) is one of the most promising technical approaches for achieving clean and sustainable fusion energy. The uniformity and structural integrity of cryogenic hydrogen isotope fuel layers directly determine implosion symmetry, hydrodynamic stability, and ultimately ignition performance \cite{lindl2004physics,caillabet2011change,kang2020unified}. In practical ICF targets, the ablator capsule and the internal fuel ice layer form a strongly coupled material system instead of two independent components. Typical candidate ablator materials include glow discharge polymer (GDP), high-density carbon (HDC) \cite{ali2018hydrodynamic,du2018recent,ho2016implosion,wang2021density}, and beryllium (Be) along with its alloys such as Be-Cu alloy \cite{kritcher2018comparison,luo2017investigation,cao2020beryllium}. These materials exhibit significant differences in atomic structure. GDP is usually amorphous, HDC is mostly nanocrystalline, and Be has a hexagonal close-packed (HCP) crystal structure.

These structural differences induce distinctly different interfacial ordering potentials on the fuel during condensation. Meanwhile, implosion instability is highly sensitive to the intrinsic surface roughness and residual microstructural inhomogeneities left during solidification \cite{haan2004design}. Therefore, the structural characteristics of the inner surface of the ablator play a crucial role in regulating the crystallization pathway, defect density, and final morphology of the cryogenic fuel layer. However, how lattice matching quantitatively dictates the growth mode, structural evolution, and crystallization quality of confined deuterium (D$_2$) has not been systematically revealed.

Under low-temperature and geometrically confined conditions, the solidification behavior of hydrogen isotopes is fundamentally different from bulk crystallization \cite{kozioziemski2010metastable,zepeda2018effect}. This difference arises from the coupled effects of spherical confinement, interfacial stress, finite-size effects, and geometric frustration. D$_2$, as a non-radioactive stable isotope, possesses unique advantages for experimental studies. At low temperatures, solid D$_2$ exhibits a HCP crystal structure. Depending on the degree of structural matching, the shell-fuel interface can either promote the coherent epitaxial growth of D$_2$ or may trigger random nucleation, grain boundary formation, and defect accumulation. Although numerous studies have improved ice layer flatness by optimizing cooling protocols \cite{lei2015solidification}, shell wettability \cite{zepeda2018effect}, and temperature uniformity \cite{lei2015solidification}, the atomic-scale mechanism by which substrate templating and lattice matching regulate confined D$_2$ crystallization remains elusive.

Previous discussions on ablator material selection have mostly focused on macroscopic properties such as opacity, density, shock wave response, and manufacturability \cite{yang2021analyzing,yang2026scaling,yan2025hydrodynamic,wang2021density,cipriani2026experimental}. They have overlooked the atomic-scale interfacial physical mechanisms that control nucleation barriers, growth modes, and structural coherence. This knowledge gap hinders the rational design of crystal templates, making it difficult to achieve the goals of suppressing polycrystallization, reducing defect generation, and lowering the inner surface roughness of fuel ice layers.

Classical nucleation theory states that a reduction in lattice mismatch between the substrate and the growing phase can lower interfacial energy and suppress defect formation \cite{kalikmanov2012classical,chernov2009single}. Solid D$_2$ is naturally stable in the HCP phase. Thus, HCP crystalline ablators such as Be and its alloys have inherent symmetry matching advantages that amorphous or nanocrystalline materials cannot match. However, how lattice matching quantitatively regulates the growth mode, structural evolution, and crystallization quality of confined D$_2$ has not been systematically clarified. Experimental efforts have been made to explore template-directed growth of high-quality hydrogen isotope ice layers \cite{kozioziemski1997crystal,shin2016supercooling,shin2018materials}. Yet, atomic-scale evolution pathways and competitive mechanisms especially under spherical confinement are difficult to obtain through direct observation. Molecular dynamics (MD) simulations have therefore become a key tool to decouple curvature-induced ordering and mismatch-induced strain.

In this work, large-scale molecular dynamics simulations based on the Feynman-Hibbs (FH) \cite{feynman2005quantum,aasen2019equation} corrected Silvera-Goldman (SG) potential function \cite{silvera1978isotropic} are employed to accurately describe the nuclear quantum effects (NQE) of D$_2$ at low temperatures \cite{jervell2025limits}. We simulate the confined crystallization process in idealized spherical HCP template shells. By self-consistently adjusting the lattice parameters, we examine the influence of geometric mismatch and highlight the role of effective template matching. Through analyses of energy evolution, structural identification, dislocation dynamics, and surface roughness, we find that there is a significant optimal lattice window for D$_2$ near 3.5 Å, which is close to the equilibrium lattice constant of HCP D$_2$. Within this window, D$_2$ grows in a coherent layer-by-layer manner, forming a near-single-crystal, low-defect ice layer with a highly smooth surface. Outside this window, crystallization shifts to island-like nucleation, accompanied by severe geometric frustration, defect accumulation, and microstructural degradation.

Experimental results from Shin et al. using the template-matching method further support the qualitative consistency of the core idea that lattice matching regulates crystal growth quality: the higher the lattice template-matching degree between the substrate and the grown crystal, the more perfect the crystal structure of the product. Taking zinc sulfide (ZnS) as an example, experimental studies show that the lattice mismatch between ZnS and H$_{2} ^{(HCP)}$ is only 0.8\%, which is the smallest among all investigated materials. With such a small lattice constant difference and high parameter matching between the ZnS substrate and H$_2$ crystal, H$_2$ can achieve high-quality epitaxial growth on the ZnS template substrate, forming a single-crystal structure with uniform crystallographic orientation and low defect density \cite{shin2016supercooling}. This experimental phenomenon acts as an external analogy for the crystallization behavior of D$_2$ in confined spherical shells revealed in this work, and corroborates that lattice matching is a universal governing factor for the growth mode and structural quality of crystalline materials, consistent with the conclusions of our confined-shell model.

This study illustrates that in ICF-relevant spherical shell systems, lattice matching rather than geometric or bulk material properties alone dominates the crystallization pathway and final quality of confined D$_2$. The relevant atomic-scale insights not only deepen the fundamental physical understanding of quantum molecular crystallization under curved confinement but also provide directly applicable design guidelines for interface engineering of next-generation high-performance ICF cryogenic targets. However, achieving such deterministic epitaxial growth​ within the curved confinement of ICF targets remains a significant challenge, requiring a deeper understanding of the interplay between lattice constraints and curvature effects.

It should be noted that D$_2$ exists in the form of quantum solids/liquids at low temperatures and exhibits significant NQE \cite{markland2018nuclear}, which therefore need to be fully considered in simulations. Although the path-integral molecular dynamics (PIMD) method can accurately describe NQE, its computational cost is excessively high for practical application in crystal growth simulations—these simulations typically involve large-scale systems (with the number of atoms ranging from hundreds of thousands to millions and a time scale of nanoseconds). To incorporate quantum effects while maintaining computational efficiency, the method of applying quantum corrections to the potential function is currently widely used, thereby realizing reasonable simulations of quantum systems within the framework of classical MD \cite{aasen2019equation,jervell2025limits}. At low temperatures, D$_2$ molecules can be approximated as spherically symmetric free rotators; thus, each D$_2$ molecule is treated as a single particle in this simulation. Although the SG potential is suitable for low-density systems, it deviates from experimental results in solid-state calculations. Therefore, FH correction was introduced in this study to improve the accuracy of describing the low-temperature properties of D$_2$ \cite{feynman2005quantum,aasen2019equation}.
\section{Models and methods}
\subsection{Model construction}
The simulations in this study focus on the crystal growth process of D$_2$ inside a spherical shell. According to classical crystal growth theory, a higher degree of lattice matching between the solute and the substrate makes it easier to achieve single-crystal growth. Hydrogen isotopes predominantly exist in the HCP phase below their triple-point temperature. The equilibrium lattice constant $a$ of D$_2$ at 10 K, calculated in this work, is approximately 3.5 Å. Accordingly, the outer spherical shell is constructed with an HCP structure, whose lattice constant $a$ ranges from 3.1 to 3.9 Å to investigate the lattice mismatch effect.

\begin{figure}
	\centering
	\includegraphics[width=0.50\textwidth]{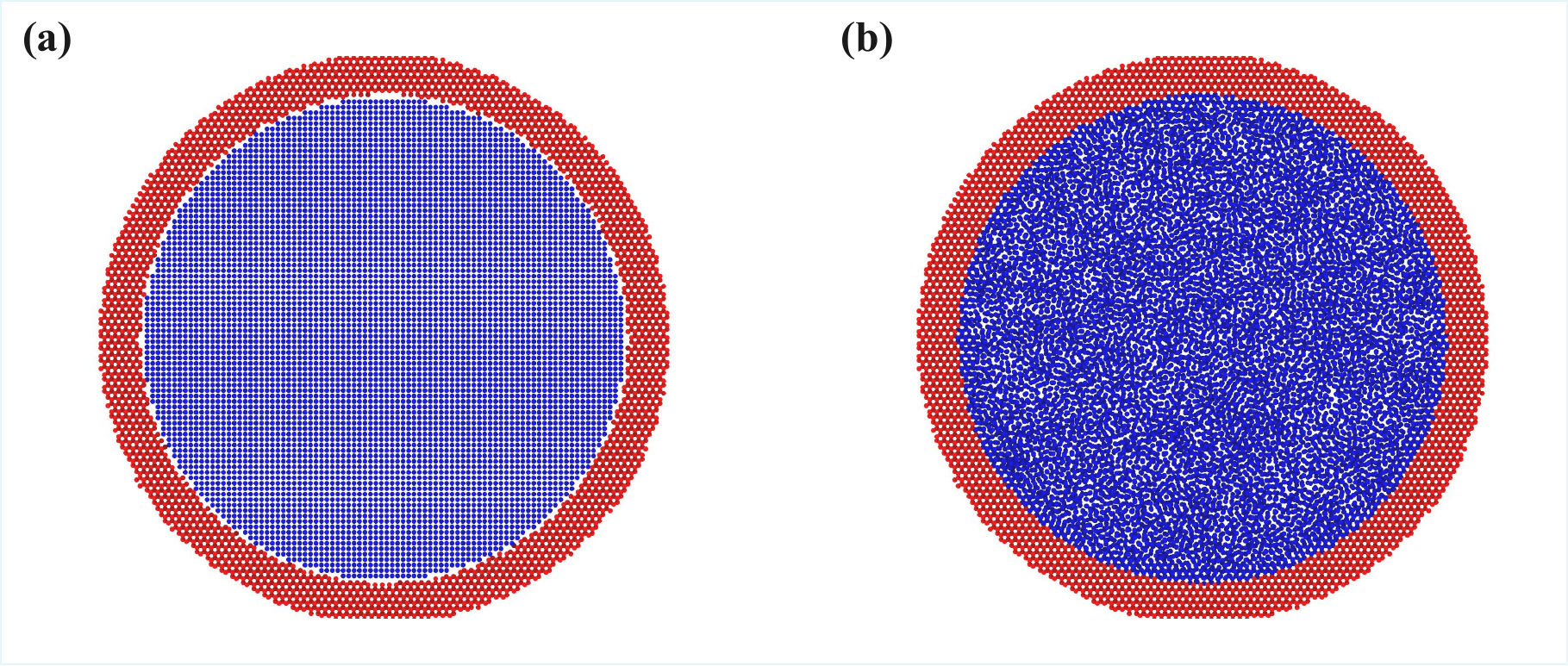} 
	\caption{(a) Initial model of the D$_2$ system; (b) Liquid D$_2$ model obtained after relaxation. Viewed along the $z$-axis; red: substrate atoms, blue: D$_2$ molecules.} 
	\label{ddmodel}
\end{figure}

The inner and outer radii of the spherical shell are set to 120 Å and 140 Å, respectively. To avoid excessive intermolecular distances, D$_2$ molecules in the inner region are packed up to a radius of 118.5 Å following a face-centered cubic (FCC) structure with a lattice constant of 5.35 Å. The entire system contains a total of 182,012 D$_2$ molecules, and the pressure at 300 K is approximately 104.15 MPa, which is close to that of ICF cryogenic targets. As illustrated in Fig.~\ref{ddmodel}(a), the atoms of the outer shell are fixed in position while retaining interactions with D$_2$ molecules, thereby acting as the substrate. Subsequently, the system is relaxed at 25 K for 1 ns using the NVT ensemble with a Nosé-Hoover thermostat (time step of 1 fs) to equilibrate the D$_2$ liquid, as shown in Fig.~\ref{ddmodel}(b).

The shell atoms are fixed to eliminate interference from substrate structural relaxation, thermal vibration, and deformation, so that we can unambiguously highlight the templating role of the substrate lattice on D$_2$ nucleation, growth pathway, and final crystal structure. This setup ensures that we accurately capture the influence of lattice matching while minimizing extraneous variables.
\subsection{Interatomic potential}
In the simulation, the intermolecular interaction potential for D$_2$ molecules adopts the SG potential \cite{silvera1978isotropic}. Considering the NQE of D$_2$ at low temperatures, the first-order Feynman-Hibbs quantum correction is applied to the SG potential \cite{jervell2025limits}, with a correction temperature of 40 K, which is consistent with Ref.~\cite{zepeda2018effect}. The corrected SG potential function can be expressed as:
\begin{equation}
	U_{FH}(r)=U_{SG}(r)+\frac{\beta\hbar^2}{24\mu}[U_{SG}^{^{\prime\prime}}(r)+2\frac{U_{SG}^{^{\prime}}(r)}{r}],
\end{equation}
where $\mu$ is the reduced mass, $\beta=(k_BT)^{-1}$, and $\hbar$ is the reduced Planck constant; $k_B$ is the Boltzmann constant and $T$ is the corrected temperature. The prime and double prime refer to the first and second derivatives, respectively, with respect to $r$.

According to the corrected SG potential function, the equilibrium lattice constant of the HCP structure for D$_2$ is determined to be 3.45 Å at 0 K and approximately 3.5 Å at 10 K. Additionally, the melting point of D$_2$ at zero pressure, obtained via the two-phase coexistence method \cite{morris1994melting,ogitsu2003melting}, is 18.1 K. The present results are in good agreement with those reported in Ref.~\cite{zepeda2018effect} and the experimental results \cite{schwalbe1984pressure}, with specific details provided in the Supporting Information, which ensures the reliability of the simulation results.

The atoms of the spherical shell are fixed in position, and only the interaction potential between the shell atoms and D$_2$ molecules is considered. This interaction is described by the Lennard-Jones (LJ) potential \cite{lenhard2024history}, whose specific form is as follows:
\begin{equation}
	U_{LJ}(r)=4\epsilon\left[\left(\frac{\sigma}{r}\right)^{12}-\left(\frac{\sigma}{r}\right)^6\right].
\end{equation}

\begin{figure}
	\centering
	\includegraphics[width=0.48\textwidth]{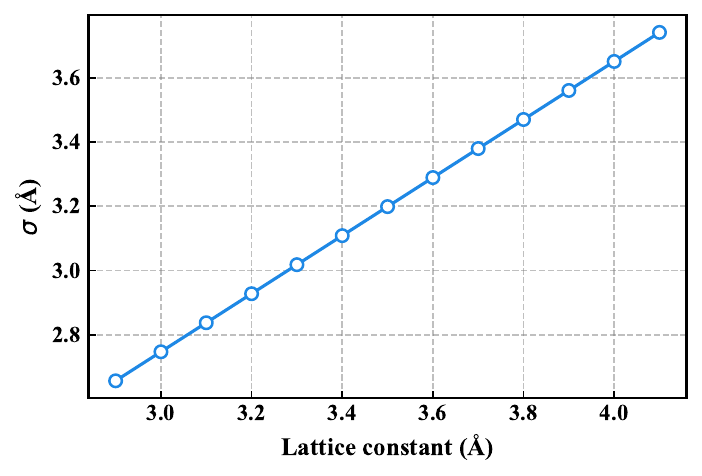} 
	\caption{The lattice constant of the spherical shell as a function of $\sigma$.} 
	\label{latsig}
\end{figure}

The parameters of the LJ potential for D$_2$ molecules are $\epsilon=2.29$ meV and $\sigma=3.07$ Å. The LJ potential parameters for the spherical shell atoms are determined as follows: its $\epsilon$ is set to 2.29 meV. Since there is a one-to-one correspondence between the equilibrium lattice constant and $\sigma$, we first set a series of $\sigma$ values, perform structural optimization to obtain the equilibrium lattice constants, and then derive the relationship between the equilibrium lattice constants and $\sigma$. Fig.~\ref{latsig} shows the relationship between the spherical shell lattice constants and $\sigma$. In the subsequent simulations, for spherical shells with different lattice constants, the corresponding $\sigma$ values are adopted as the LJ potential parameters of the spherical shell atoms.

\subsection{Molecular dynamics}
All simulations were performed using the Large-scale Atomic/Molecular Massively Parallel Simulator (LAMMPS) MD code \cite{thompson2022lammps}. After obtaining the liquid D$_2$ model, the model was subjected to a cooling simulation under the NVT ensemble, with temperature controlled by a Nosé-Hoover thermostat. The cooling process was conducted at a rate of $1.5\times10^9$ K/s from 25 K to 10 K, with a total duration of 10 ns. During this process, the phase transition of D$_2$ from liquid to solid and subsequent complete solidification was observed. Finally, a relaxation simulation was performed at 10 K for 1 ns to characterize the inner surface roughness of spherical shells with different lattice constants. To eliminate numerical errors, five independent simulations were conducted for each case.

It should be noted that the cooling rate of $1.5\times10^9$ K/s employed here is significantly higher than those achievable in macro-scale experiments. This choice is necessitated by the inherent time-scale limitations of MD simulations. However, this rate is sufficiently moderate within the atomic scale to allow for structural relaxation and the capture of essential liquid-to-solid phase transition dynamics, providing a reliable theoretical basis for the analysis of crystallization behavior.
\section{Results and discussion}
\subsection{The crystal growth process of D$_2$}
To conduct an in-depth study of the crystal growth of D$_2$ inside spherical shells with different lattice parameters, we performed a detailed analysis of the energy variations, growth modes, and other relevant characteristics of D$_2$ during the crystallization process.

\begin{figure}
	\centering
	\includegraphics[width=0.50\textwidth]{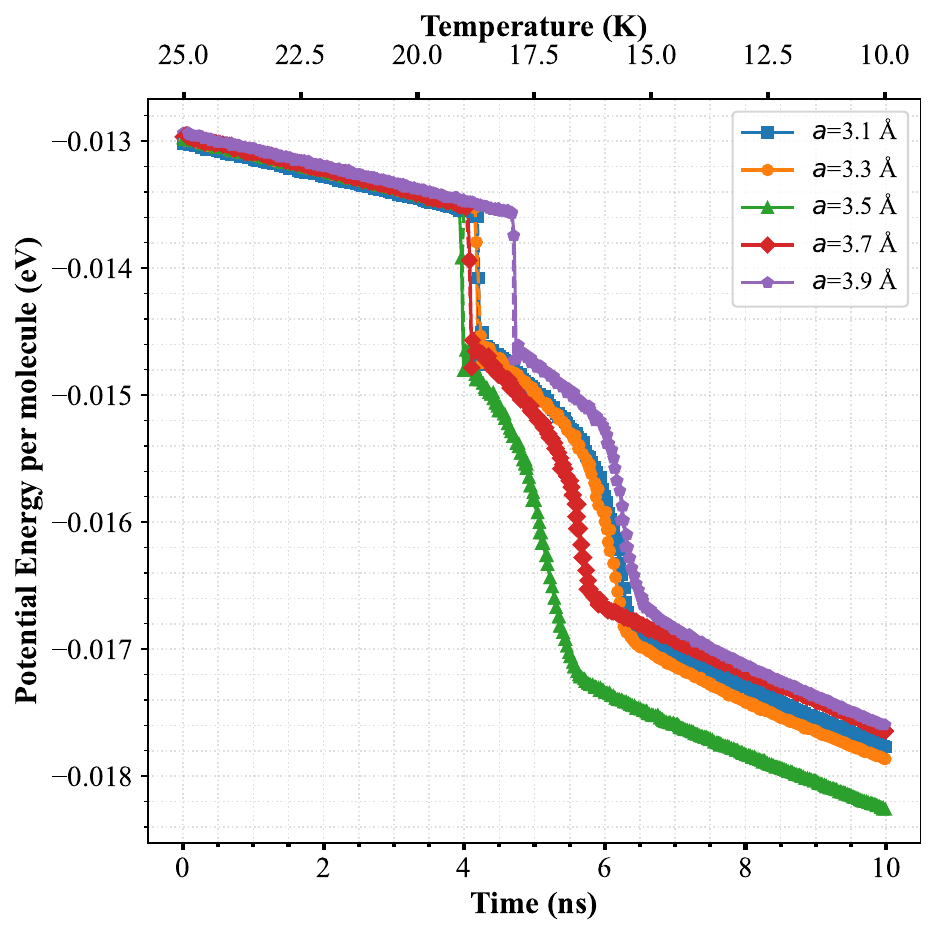} 
	\caption{Single-molecule potential energy of D$_2$ as a function of temperature (time) for spherical shell substrates with different lattice constants.} 
	\label{energytemptime}
\end{figure}

First, we analyzed the variation of the single-molecule potential energy of D$_2$ with temperature (and time) on spherical shell substrates of different lattice constants. Fig.~\ref{energytemptime} shows the single-molecule potential energy–temperature (time) curves of D$_2$ corresponding to substrates with different lattice constants. The results indicate that, as the temperature gradually decreases to a critical value, D$_2$ exhibits a prefreezing phenomenon accompanied by an abrupt change in the single-molecule potential energy. This abrupt change signifies the onset of the solid–liquid coexistence stage, during which D$_2$ transforms from liquid to solid until it is fully solidified.

We can further observe that during the initial cooling stage, the single-molecule potential energy of D$_2$ on substrates with different lattice constants is relatively similar. Subsequently, D$_2$ grown on the substrate with a lattice constant of 3.5 Å exhibits the prefreezing phenomenon first and enters the solid-liquid coexistence stage earlier than D$_2$ grown on other substrates. Moreover, after the formation of a complete solid, D$_2$ grown on the substrate with a lattice constant of 3.5 Å has a lower energy than that on other substrates.

In addition, the smaller the lattice mismatch, the higher the critical temperature at which D$_2$ transitions from the liquid state to the solid-liquid coexistence stage. Meanwhile, the system remains in a lower energy state than that on other substrates, which indicates that under the same conditions, the lattice matching degree between the substrate and D$_2$ exerts a significant influence on the energy state of the system.

Interestingly, in contrast to the findings reported by Lei et al., they found that an increase in the cooling rate induces solidification at a lower temperature; that is, a higher cooling rate results in a lower onset temperature of solid-liquid coexistence, with a corresponding rise in the degree of supercooling \cite{lei2015solidification}. In the present study, however, under the condition of a fixed cooling rate, we found that a smaller lattice mismatch degree between the substrate and D$_2$ molecules yields a higher critical temperature for the transition of D$_2$ from the liquid state to the solid-liquid coexistence state. These findings indicate that, in addition to the cooling rate, the degree of lattice matching between the substrate and D$_2$ molecules is also an important parameter governing the solidification behavior, and both factors jointly modulate the onset temperature of solidification and the phase transition process.

To gain in-depth insights into the microscopic evolution mechanism of the three stages of D$_2$ crystallization and growth within closed spherical shells, this study conducts analysis at the atomic scale. Drawing on the classification approach proposed in Ref.~\cite{zhan2023multiple}, the D$_2$ molecules in the system are divided into two categories: precursor particles (abbreviated as prec) and geometrically frustrated particles. Among these, within the category of geometrically frustrated particles, the focus is placed on tetrahedral aggregates (abbreviated as teag) with characteristic topological structures, which serve as the core to analyze the structural transformation laws during the crystallization process.

To identify crystalline structures, we utilize the bond order parameters proposed by Steinhardt et al \cite{steinhardt1983bond,tan2014visualizing,rein1996numerical,lechner2008accurate,mickel2013shortcomings,arai2017surface}. The detailed procedure is as follows:

For particle $i$, we first determine its neighboring particles within a distance $r_s$, which corresponds to the first minimum of the radial distribution function. We employ the neighbor-correction method introduced in Ref.~\cite{zhan2023multiple} to count the number of neighboring particles $N_b(i)$. The basic order parameter is defined as:
\begin{equation}
	q_{l,m}(i) = \frac{1}{N_b(i)} \sum_{j=1}^{N_b(i)} Y_{l,m}(\theta_{ij}, \phi_{ij}),
\end{equation}
where $Y_{l,m}(\theta_{ij}, \phi_{ij})$ denotes the spherical harmonic function with $m \in [-l,l]$, and $\theta_{ij}$ and $\phi_{ij}$ are the polar and azimuthal angles of the vector $\vec{r}_{ij}$ pointing from particle $i$ to its neighboring particle $j$, respectively.

We further coarse-grain $q_{l,m}(i)$ to obtain:
\begin{equation}
	\bar{q}_{l,m}(i) = \frac{1}{N_c(i) + 1} \sum_{j=0}^{N_c(i)} q_{l,m}(j),
\end{equation}
here, $N_c(i) \leq N_b(i)$ represents the number of neighboring particles that share the same phase (liquid-like or solid-like) as particle $i$. The summation includes all these $N_c(i)$ neighboring particles as well as particle $i$ itself (denoted by $j=0$).

Accordingly, the bond orientational order parameter of particle $i$ is expressed as:
\begin{equation}
	q_l(i) = \left[ \frac{4\pi}{2l + 1} \sum_{m=-l}^l \left| \bar{q}_{l,m}(i) \right|^2 \right]^{1/2}.
\end{equation}

Crystalline particles and precursors are identified by the criteria $q_6>0.4$ and $0.27< q_6<0.4$, respectively. Particles with $q_6<0.27$ and a number of neighboring particles greater than or equal to 12, and an association with at least 12 adjacent quasiregular tetrahedra are defined as geometrically frustrated particles.

\begin{figure}
	\centering
	\includegraphics[width=0.50\textwidth]{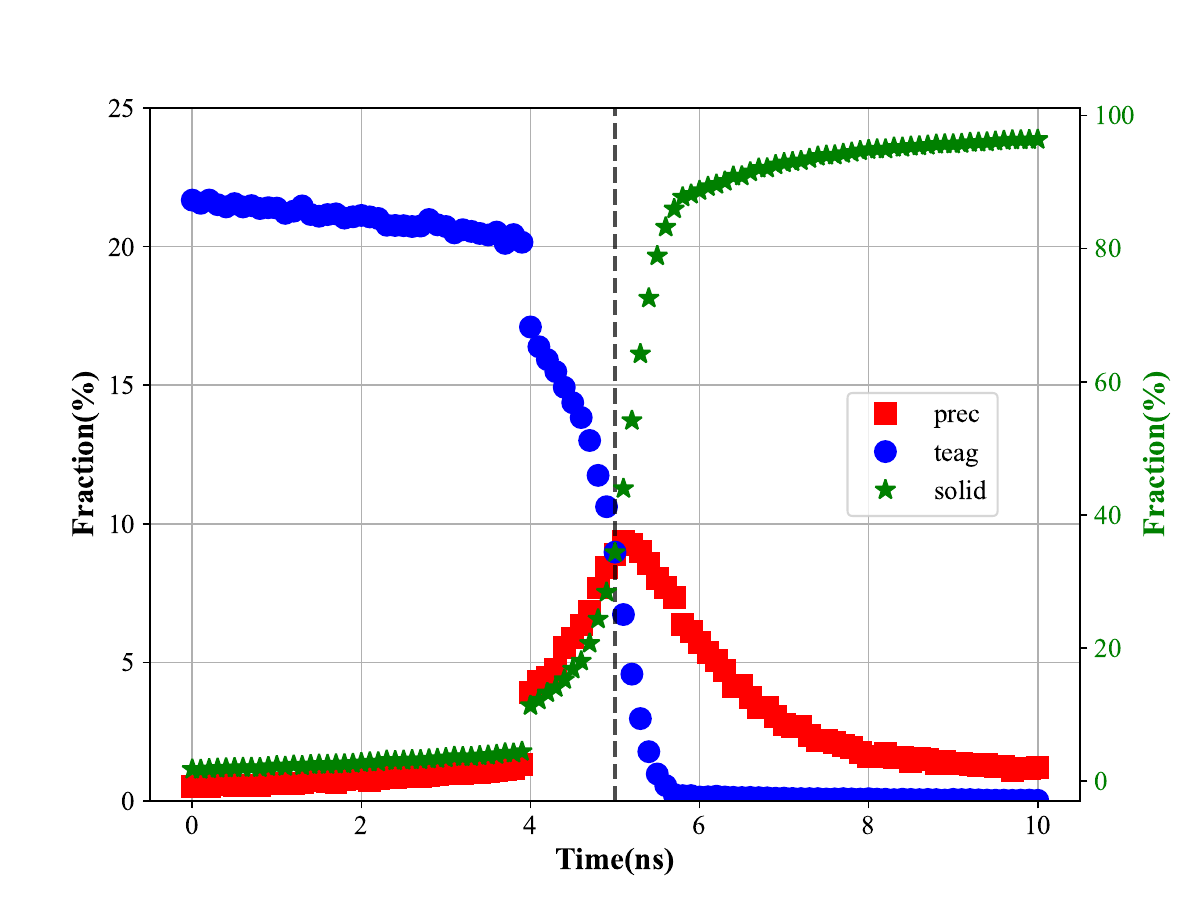} 
	\caption{Variation of the proportions of precursor particles (prec), geometrically frustrated particles (teag), and the solid phase with time during the crystallization process when the substrate lattice constant is 3.5 Å, where the green y-axis on the right specifically denotes the proportion of the solid phase (the dashed line represents 5 ns).} 
	\label{farcteagprec}
\end{figure}

Fig.~\ref{farcteagprec} shows the evolution of the proportions of precursor particles (prec), geometrically frustrated particles (teag), and the solid phase over time during the crystallization process when the substrate lattice constant is set to 3.5 Å. 

When combined with Fig.~\ref{energytemptime}, it can be observed that during the initial cooling stage, the proportion of precursor particles (prec) increases very slowly while the proportion of geometrically frustrated particles (teag) decreases equally slowly; this stage can thus be regarded as a competitive phase between precursors and geometrically frustrated particles. Subsequently, at approximately 4 ns, D$_2$ initiates the liquid-solid phase transition and enters the solid-liquid coexistence stage, where the proportion of precursors rises rapidly and the proportion of geometrically frustrated particles drops sharply. Finally, around 6 ns, the liquid-solid phase transition is completed, with only a small fraction of precursors remaining in the process of transforming into the solid phase. Notably, at approximately 5 ns, the proportion of precursors reaches its maximum, which exactly corresponds to the most rapid liquid-solid transition and the fastest decline in the proportion of geometrically frustrated particles.

Ostwald's stepwise nucleation theory states that when a system in an unstable or metastable state transitions to a stable state, the system tends to first reach a metastable state whose free energy is closest to that of the initial state, rather than directly forming the stable state with the lowest free energy \cite{ten1999homogeneous,schmelzer2017crystals}. According to this theory, during the crystallization and growth of D$_2$, particles first aggregate to form metastable precursors. Our above conclusion that the maximum proportion of precursor particles precisely corresponds to the most rapid liquid-solid phase transition rate also indirectly confirms this point.

\begin{figure}
	\centering
	\includegraphics[width=0.50\textwidth]{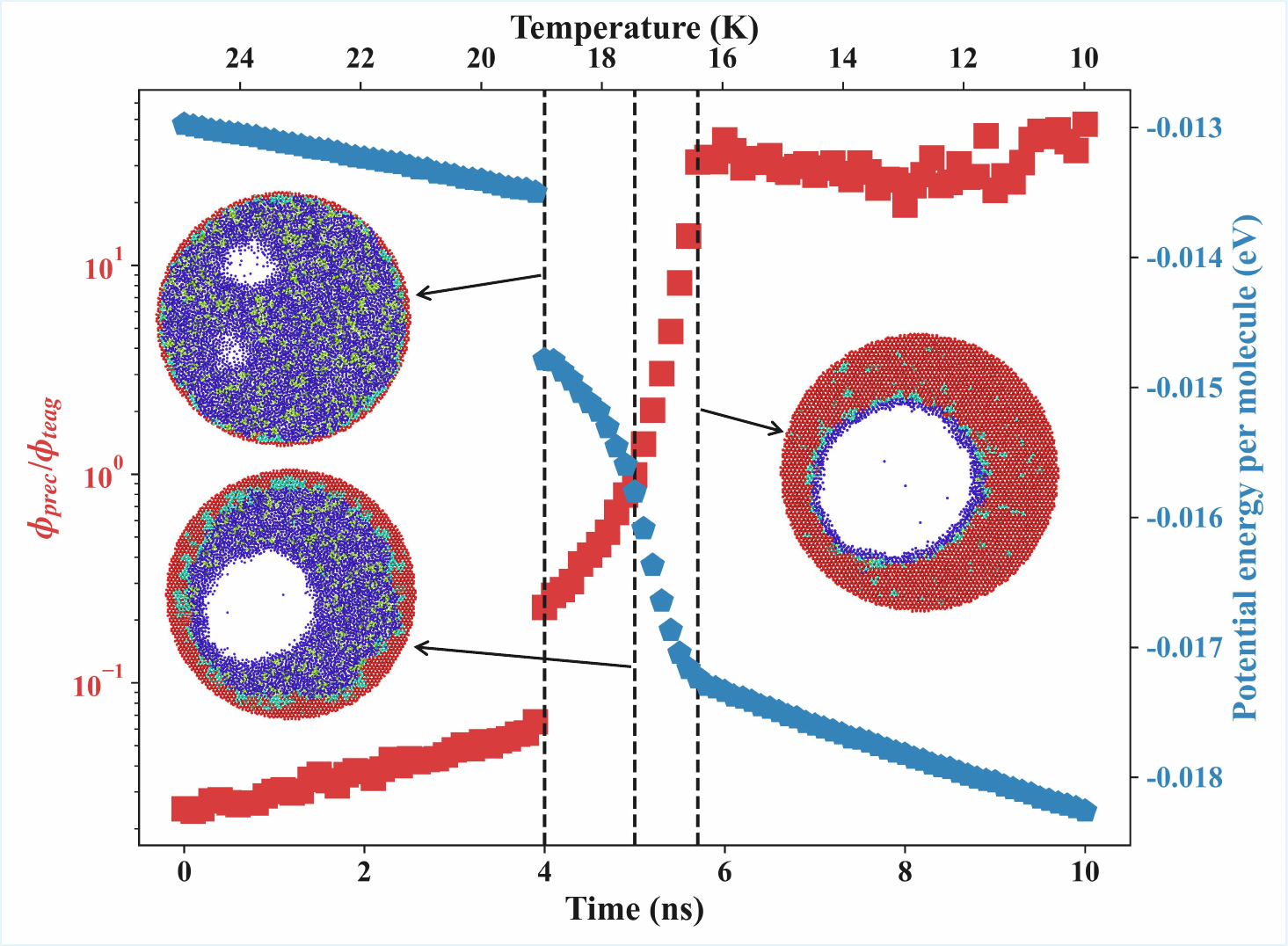} 
	\caption{When the substrate lattice constant is 3.5 Å, this figure shows the ratio of the precursor particle fraction ($\phi_{prec}$) to the geometrically frustrated particle fraction ($\phi_{teag}$) as a function of time. For better comparison, the curve of single-molecule potential energy versus time is also included (displayed as the blue curve). In the crystal growth schematic, red represents solid D$_2$, cyan on the solid surface denotes precursor particles, green indicates geometrically frustrated particles, and blue stands for other liquid D$_2$.} 
	\label{precteagfrac}
\end{figure}

Fig.~\ref{precteagfrac} presents the time evolution of the ratio between the precursor particle fraction ($\phi_{prec}$) and the geometrically frustrated particle fraction ($\phi_{teag}$) when the substrate lattice constant is 3.5 Å. Three distinct stages of D$_2$ crystal growth can be clearly identified from the figure. During the liquid-solid phase transition, nucleation first initiates on the substrate surface, and precursor particles gradually transform into solid-state particles—this observation is further supported by the structural schematics at 4 ns and 5 ns. Epitaxial growth is achieved exclusively at 3.5 Å, proceeding in a coherent layer-by-layer manner consistent with Ostwald’s stepwise nucleation theory, ultimately forming near-single-crystal structures with minimal defects.​ In contrast, any deviation from this lattice constant suppresses epitaxy, leading to island-like nucleation and polycrystalline structures. This growth pattern is in excellent agreement with the stepwise nucleation theory described earlier in this work \cite{lozovoy2020kinetics,velikov2002layer,ten1999homogeneous,schmelzer2017crystals}.

\begin{figure*}
	\centering
	\includegraphics[width=1.0\textwidth]{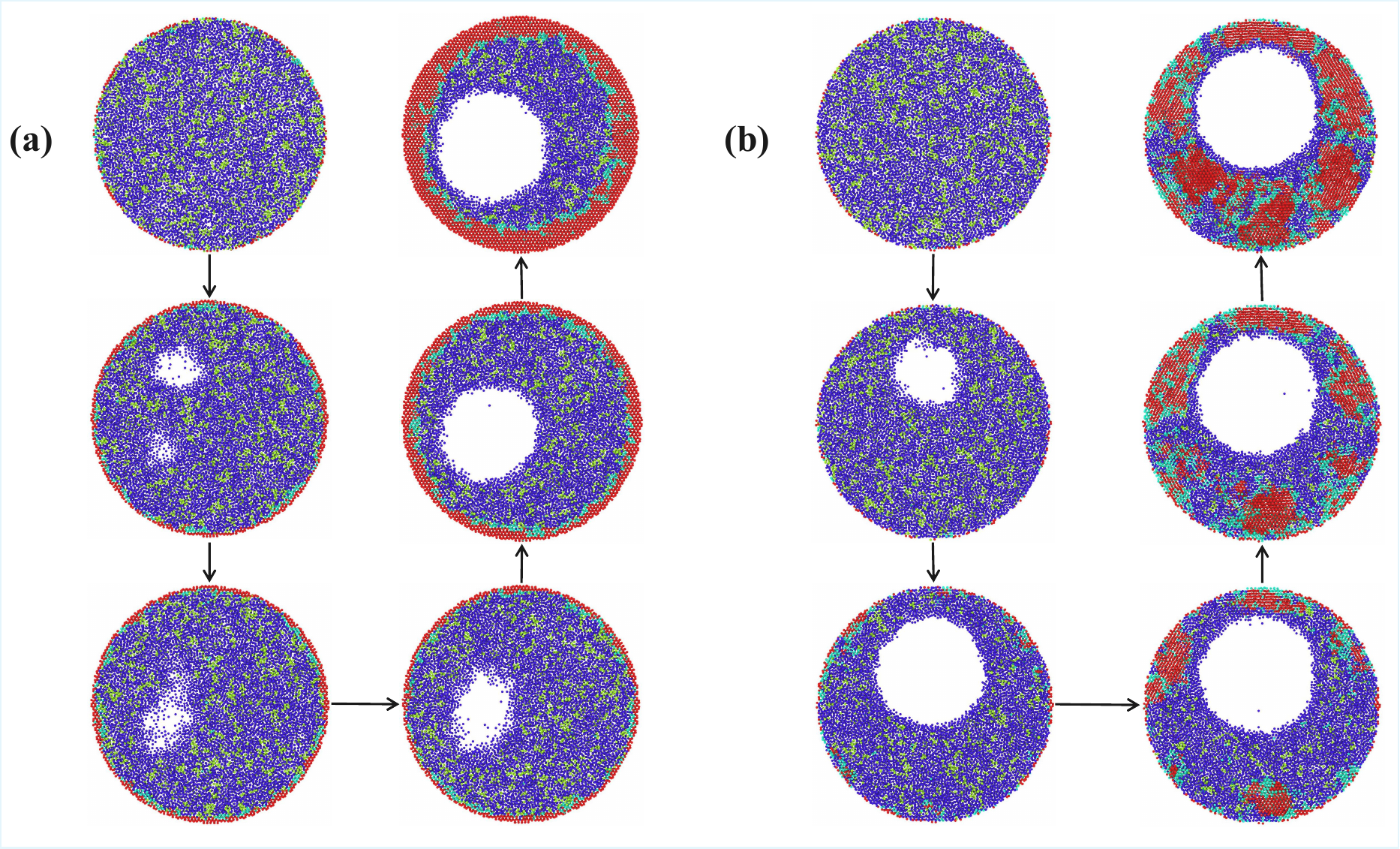} 
	\caption{Schematic diagrams of the crystallization growth process of D$_2$ in substrates with different lattice constants, where panel (a) corresponds to a substrate with a lattice constant of 3.5 Å and panel (b) to one with 3.9 Å (In the crystal growth schematic, red represents solid D$_2$, cyan on the solid surface denotes precursor particles, green indicates geometrically frustrated particles, and blue stands for other liquid D$_2$). Viewed along the $z$-axis with the substrate removed.} 
	\label{growthmech}
\end{figure*}

In contrast, when a different substrate is employed and the lattice mismatch increases, the crystal growth process of D$_2$ no longer follows the aforementioned layered growth mechanism, but instead transitions to an island-like growth mode \cite{puurunen2004island}. Figs.~\ref{growthmech}(a) and (b) show the crystallization growth process of D$_2$ when the substrate lattice constants are 3.5 Å and 3.9 Å, respectively. It can be observed that the two exhibit completely different crystallization growth mechanisms: when the substrate lattice constant is 3.9 Å, the growth of D$_2$ tends to undergo local nucleation, forming numerous sub-crystals, and crystal growth is achieved through the expansion of these sub-crystals—rather than the layer-by-layer growth we described earlier. Furthermore, when other substrates are used (with lattice constants of 3.1 Å, 3.3 Å, and 3.7 Å), similar island-like growth modes can also be observed. Relevant images and trajectory animations are available in the supporting information.

Notably, our results exhibit a certain discrepancy from those reported in Ref.~\cite{zhan2023multiple}. In their work, the maximum value of the ratio between the precursor particle fraction and the geometrically frustrated particle fraction ($\phi_{prec}/\phi_{teag}$) corresponds to the fastest nucleation rate. In contrast, our findings indicate that the peak growth rate of this ratio is correlated with the maximum nucleation rate, whereas the attainment of its maximum value coincides with the completion of the liquid–solid phase transition. We attribute these discrepancies to the fact that our simulations focus on crystallization growth under confined conditions, whereas their theoretical model is developed for supercooled liquids in open systems. Nevertheless, we can confirm that $\phi_{prec}/\phi_{teag}$ plays a highly important role in the crystallization growth process.

Moreover, the modulation of the D$_2$ crystal growth mode by the lattice mismatch of the inner substrate in this study shares a profound physical connection with the wetting effect in confined environments reported in Ref.~\cite{zepeda2018effect}. Both mechanisms govern nucleation sites and growth pathways by modulating the interfacial interaction energy, leading to highly parallel spectra of experimental phenomena: low lattice mismatch (high structural compatibility) and complete wetting (strong D$_2$-wall attraction) correspond to a low interfacial energy state, promoting heterogeneous nucleation and two-dimensional layer-by-layer growth, ultimately yielding large-grained, low-defect, and densely structured crystals; moderate mismatch/partial wetting results in increased interfacial energy and discrete nucleation sites, manifesting as three-dimensional island growth and grain coalescence; high mismatch/non-wetting further suppresses wall nucleation, instead inducing bulk homogeneous nucleation and forming numerous quasi-spherical grains with the highest defect density.

However, there are key differences in their physical essence and constraints: the wetting effect in Ref.~\cite{zepeda2018effect} essentially modulates the strength of van der Waals interactions, directly determining the adsorption intensity of D$_2$ molecules on the wall and the nucleation driving force, with defects mainly arising from grain boundaries during polycrystalline coalescence. In contrast, the lattice mismatch in this study represents structural compatibility modulation dominated by elastic strain energy, where mismatch strain between the substrate and crystal lattice affects nucleation orientation and stress relaxation pathways, with defects primarily taking the form of dislocations, stacking faults, and other strain-relief mechanisms. Notably, under the constraint of spherical shell curvature, even a small lattice mismatch may induce significant stress gradients due to curvature effects, further influencing defect distribution and growth kinetics. Therefore, substrate engineering for high-quality D$_2$ crystal growth must jointly consider the dual influences of chemical interactions (wettability) and structural matching (lattice constant).

The differentiation of D$_2$ growth mechanisms dominated by the substrate lattice mismatch degree ultimately manifests as differences in crystallization quality. Our subsequent crystallization quality analysis has also confirmed that layered growth tends to facilitate the growth of single crystals, whereas island-like growth is prone to result in the growth of polycrystalline structures; furthermore, characterizations of dislocation lines and associated defects corroborate the aforementioned discussions. This finding therefore underscores the critical regulatory role of the lattice mismatch between the substrate and D$_2$ in governing the growth pathways of D$_2$ crystals.

\subsection{Discussion on Crystal Growth Modes}
\begin{figure}
	\centering
	\includegraphics[width=0.50\textwidth]{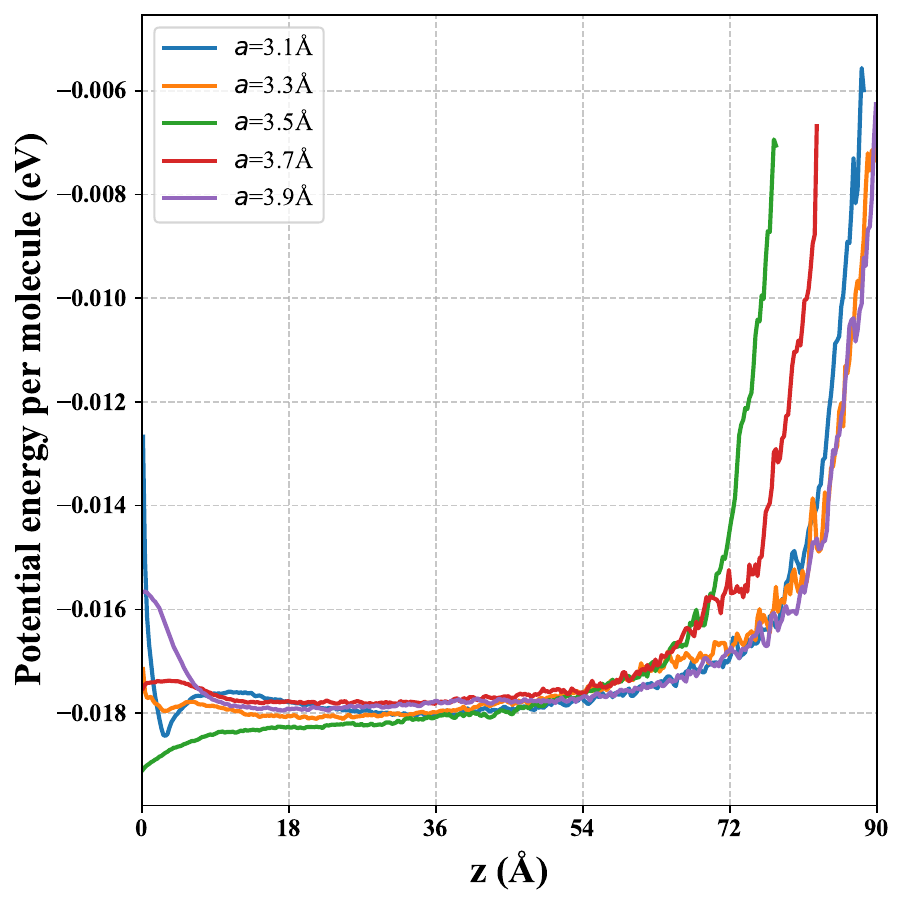} 
	\caption{Radial single-molecule potential energy of D$_2$ grown on different substrates after complete crystallization (z is the distance from the substrate spherical shell).} 
	\label{rdpe}
\end{figure}

To reveal the physical mechanism underlying the distinct growth modes of D$_2$ crystals dictated by lattice matching, we computed the radial single-molecule potential energy after complete crystallization, as shown in Fig.~\ref{rdpe}. For the substrate with a lattice constant of 3.5 Å, the radial potential energy increases monotonically from the shell interface inward. In contrast, for  D$_2$ grown on lattice-mismatched substrates, the radial single-molecule potential energy first decreases to a local minimum, remains nearly flat, and then rises abruptly.

This difference in potential energy evolution essentially corresponds to the dominant distinction between the curvature effect and the stress induced by interfacial lattice mismatch. Specifically, the lattice-matched condition of 3.5 Å corresponds to an elastic regime, where the single-molecule potential energy increases linearly and monotonically from the spherical shell interface toward the core. This potential energy gradient directly confirms that the D$_2$ crystals grown under this condition form a coherent single-crystalline structure. The continuous rise in potential energy is essentially the elastic work done by the lattice to adapt to the intrinsic curvature of the spherical shell, and no significant lattice mismatch stress is present in this process, indicating that the energy regulation of crystal growth is dominated by the curvature effect.

In contrast, when the lattice constants are greater than or equal to 3.7 Å, the system is in a plastic regime dominated by tensile strain: the interfacial tensile stress exceeds the elastic bearing capacity of the lattice and cannot be relieved by elastic deformation, but is instead released through plastic relaxation mechanisms such as the formation of misfit dislocations and grain boundaries, ultimately leading to the formation of a discontinuous polycrystalline structure.

As for lattice constants less than or equal to 3.3 Å, the system falls into a jammed regime dominated by compressive strain: molecular overcrowding caused by lattice mismatch traps the system in a frustrated metastable state, which not only hinders coherent elastic growth but also precludes effective plastic relaxation, ultimately resulting in distorted crystal growth.

For D$_2$ grown on all substrates, the radial single-molecule potential energy increases sharply at a radial distance of approximately 54 Å from the spherical shell. We attribute this steep rise to the region near the inner surface of the ice layer, where molecules are under-coordinated. The significantly lower coordination number relative to molecules in the interior gives rise to a higher single-molecule potential energy, resulting in the rapid surge observed in the potential energy profile.

It can be further observed that after crystallization, the D$_2$ layer grown on the 3.5 Å substrate exhibits a smaller radial thickness and the radial potential energy curve reaches the surface region more rapidly, indicating a more compact structure. In contrast, the D$_2$ layers formed on substrates with larger lattice mismatches contain more defects and pores, resulting in a larger radial thickness and a delayed rise in the potential energy curve, which confirms their poorer compactness and structural integrity.

Subsequently, under the spherical-symmetric approximation, we calculated the radial stress $\sigma_{rr}$ and the hoop stress $\sigma_{\theta\theta}$ for each molecule. First, the Cartesian components of the atomic stress tensor $\sigma_{xx}, \sigma_{yy}, \sigma_{zz}, \sigma_{xy}, \sigma_{xz}, \sigma_{yz}$ (in units of GPa) were obtained from LAMMPS. With the radial unit vector defined as $\mathbf{n}=(x/r, y/r, z/r)$, the radial stress is given by:
\begin{equation}
	\sigma_{r} = \mathbf{n}^T \sigma \mathbf{n}.
\end{equation}

Under the condition of spherical symmetry, the hoop stress can be derived from the trace of the stress tensor $\operatorname{Tr}(\boldsymbol{\sigma})$ and the radial stress:
\begin{equation}
	\sigma_{hoop} = \frac{\operatorname{Tr}(\boldsymbol{\sigma}) - \sigma_{r}}{2}.
\end{equation}

\begin{figure*}
	\centering
	\includegraphics[width=1.0\textwidth]{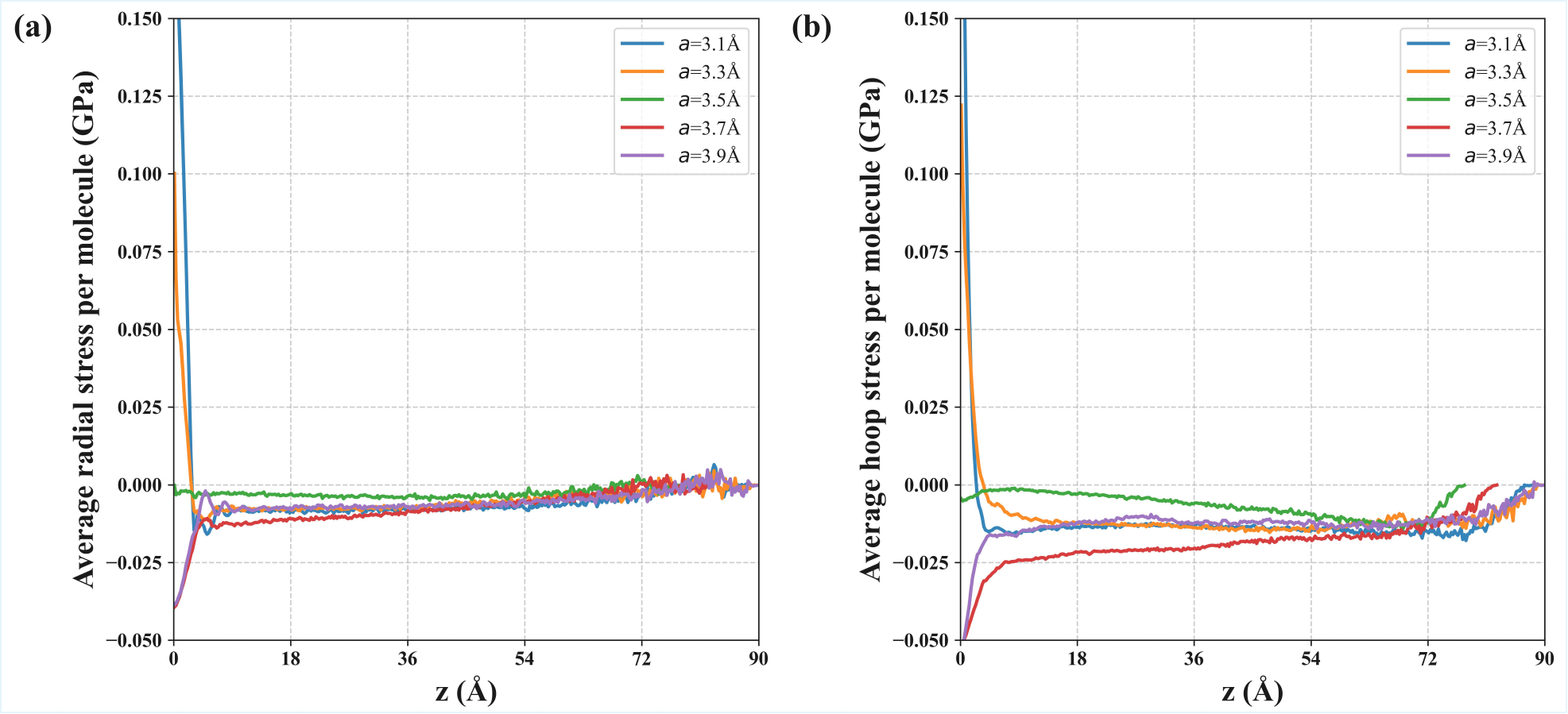} 
	\caption{Average (a) radial and (b) hoop single-molecule stresses of D$_2$ grown on various substrates after full crystallization as a function of distance z from the substrate spherical shell.} 
	\label{stress}
\end{figure*}

Fig.~\ref{stress} shows the radial and hoop stresses as functions of the radial distance from the spherical shell after complete crystallization. It can be seen that the interfacial stress induced by lattice mismatch is characterized by strong localization and short-range effects. Both radial and hoop stresses (whether tensile or compressive) decay rapidly to negligible levels within approximately 10 Å from the spherical-shell interface, corresponding to only 2–3 layers of solid molecules.

This quantitative result suggests that the interfacial stress does not act as a long-range field governing the entire growth process, but instead serves as an initial trigger: within the first few molecular layers adjacent to the substrate, it determines the type of stress (tensile or compressive) and immediately nucleates the corresponding microstructural defects. Beyond this thin interfacial region, subsequent crystal growth proceeds under conditions where the external stress field has been largely released, and the growth behavior is dominated by defects initiated at the interface, such as misfit dislocations, grain boundaries, or jammed clusters.

\begin{figure}
	\centering
	\includegraphics[width=0.50\textwidth]{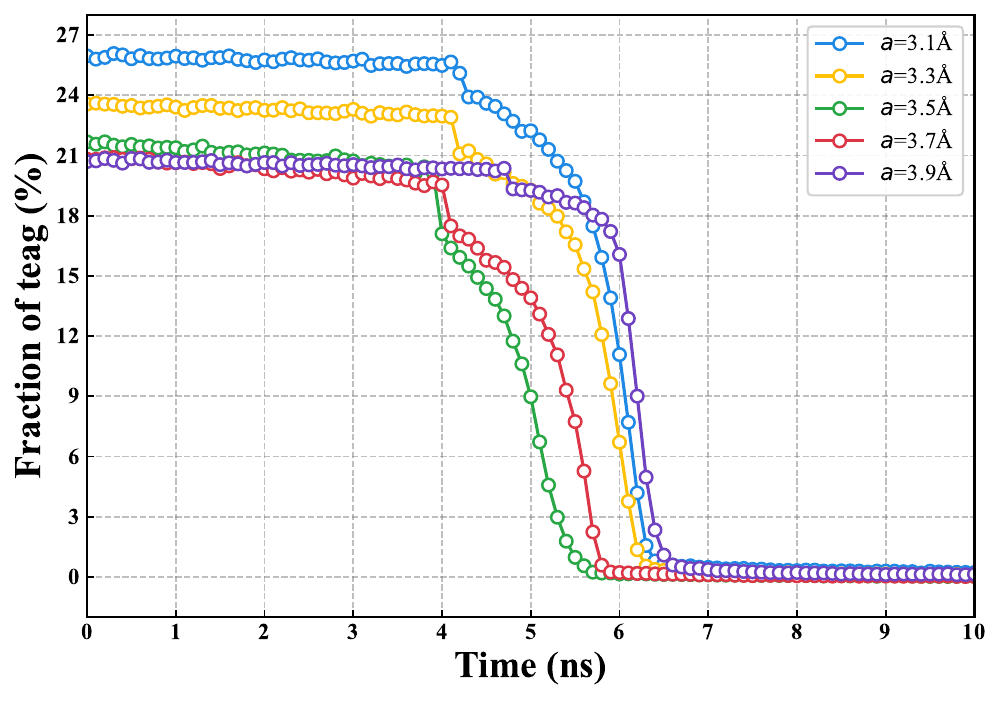}
	\caption{Geometrically frustrated particle fraction of D$_2$ grown on different substrates as a function of time.} 
	\label{newteag}
\end{figure}

To further validate the above conclusions and correlate the short-range interfacial stress with the long-range structural evolution, we statistically analyzed the time evolution of the fraction of geometrically frustrated motifs in D$_2$ grown on substrates with different lattice constants, as shown in Fig.~\ref{newteag}. Evidently, compressive strain (3.1 Å) increases the population of frustrated clusters during the pre-nucleation stage. In this case, a large number of molecules aggregate on the spherical shell surface, and the high molecular packing density facilitates the formation of tetrahedral assemblies, thereby promoting the generation of geometric frustration. In comparison, tensile strain (3.9 Å) primarily induces a kinetic delay in the relaxation of the frustrated state. A comparative analysis between the lattice-matched substrate (3.5 Å) and the mismatched one (3.9 Å) provides clear insight: despite similar initial degrees of geometric frustration, the onset of crystallization on the 3.9 Å substrate is delayed by approximately 1.5 ns. This hysteresis demonstrates that lattice mismatch effectively raises the free-energy barrier for the liquid-solid phase transition, confining D$_2$ in a metastable state stabilized by geometric frustration.

By combining the short-range nature of interfacial stress and the dynamic evolution of frustrated clusters, a two-stage growth mechanism is suggested, namely short-range stress triggering and long-range defect-dominated growth. In the initial stage, short-range stress appears to determine the molecular packing manner, where compressive stress leads to molecular crowding and tensile stress induces lattice expansion. In the subsequent growth stage, the external stress field decays rapidly, and the system evolves under the control of defects nucleated at the interface. For tensile lattice mismatch, misfit dislocations formed at the interface continuously accommodate strain, which delays the crystallization kinetics and eventually results in polycrystalline island structures. For compressive mismatch, initially crowded clusters are kinetically trapped without sufficient structural relaxation, leading to jammed and distorted crystals. The distinct roles of stress as the trigger and defects as the carrier of initial conditions are consistent with why a stress field effective only within 10 Å can still exert a key influence on the structure of crystals grown to more than 70 Å in thickness.

Overall, lattice mismatch regulates the growth mode of D$_2$ crystals through a two-stage mechanism: short-range stress triggering and long-range defect-dominated growth. At the lattice-matched condition of 3.5 Å, the curvature effect of the spherical shell substrate appears to dominate, facilitating the formation of a coherent single-crystalline structure via a layered growth mode. In contrast, any deviation from this lattice-matched condition gives rise to either compressive or tensile stress at the interface, where the stress effect becomes predominant and suppresses the regulatory role of the curvature effect in crystal growth. Specifically, tensile stress triggers plastic relaxation and relieves stress through the generation of misfit dislocations and grain boundaries, whereas compressive stress causes molecular overcrowding, trapping the system in a frustrated metastable state and precluding effective relaxation. Such interfacial stress not only converts the growth mode of D$_2$ crystals to the island-like mode in all cases but also elevates the free-energy barrier for the liquid-solid phase transition, induces plastic relaxation, causes kinetic delay or even inhibition of the relaxation process of the frustrated state, ultimately leading to the formation of polycrystalline or distorted crystal structures.

It should be noted that in this study, attempts were made to directly extract and analyze the interfacial stress, particularly to visually characterize the stress distribution features via interfacial stress contour plots. However, during the actual analysis, considerable difficulties were encountered in the accurate interpretation of the stress contour plots, making it challenging to effectively capture the key information of the stress distribution. Consequently, in the atomic-scale simulations, we did not observe a clear systematic correlation between the stress distribution and the final microstructure from the stress contour plots.

This result is consistent with the widely accepted consensus in the field: the accurate calculation and analysis of local stress at the atomic scale are inherently challenging, as the definition of stress is highly sensitive to the choice of averaging volume, stress decomposition schemes, and thermal fluctuations \cite{zhou2003new,chen2016origin,subramaniyan2008continuum,xu2009investigation,liu2009compute,dai2021comparative}.

In view of this, we adjusted our analytical strategy and focused instead on the extraction and analysis of radial and hoop stresses. Using these two characteristic stress components, we indirectly infer the distribution law of interfacial stress and its regulatory effect on the growth mode of D$_2$ crystals, thereby supporting the above analysis.

\subsection{The analysis of D$_2$ crystallization quality}
\begin{figure}
	\centering
	\includegraphics[width=0.50\textwidth]{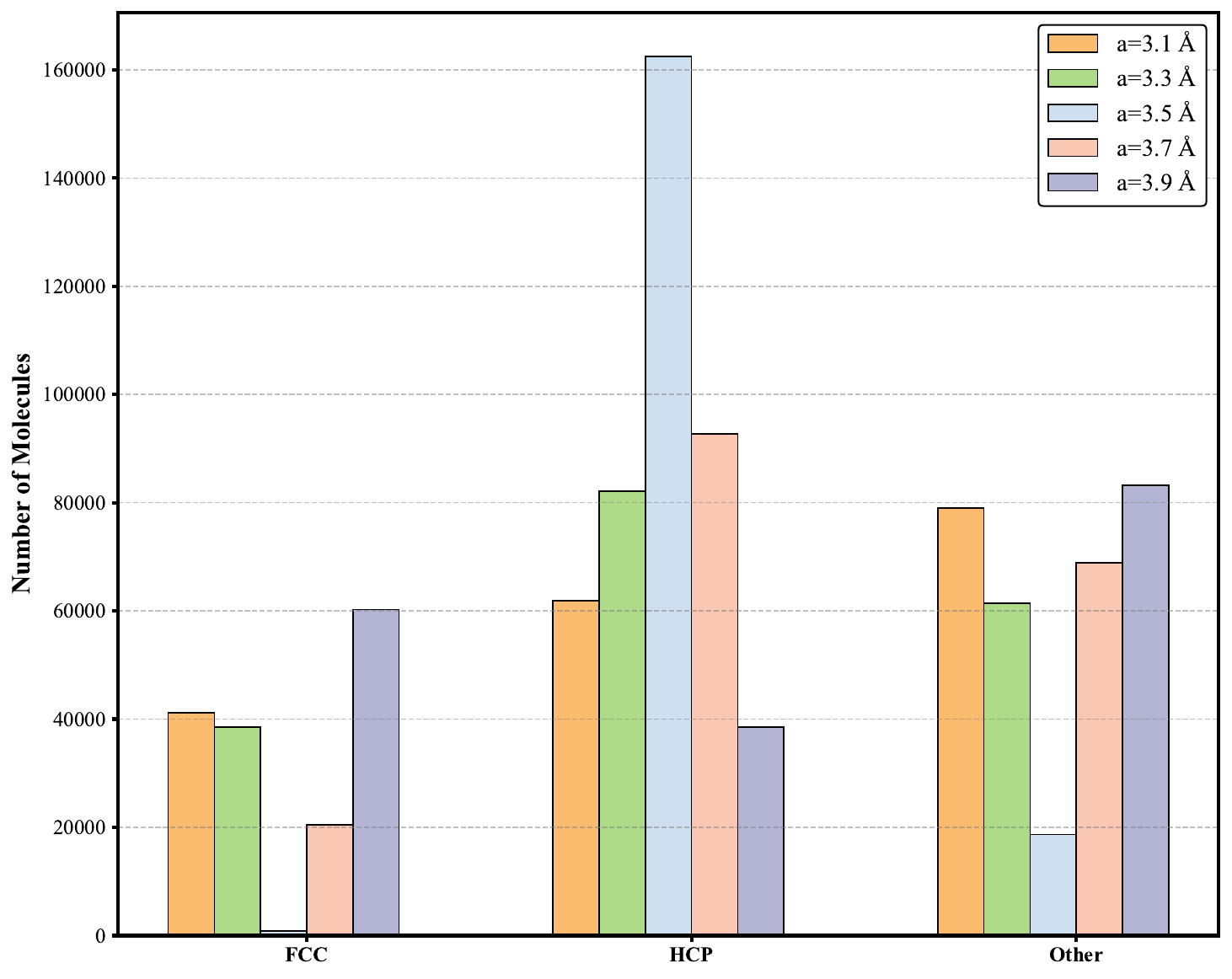} 
	\caption{Types and molecular counts of D$_2$ molecules under substrates with different lattice constants.} 
	\label{stat}
\end{figure}

After the crystallization was completed, we statistically analyzed the structural types of D$_2$ molecules in substrates with different lattice constants, and the results are presented in Fig.~\ref{stat}. It can be seen that the D$_2$ molecules in the substrate with a lattice constant of 3.5 Å are predominantly in the HCP structure, while a small number of surface molecules are classified as other types via the Polyhedral Template Matching (PTM) method \cite{larsen2016robust,stukowski2009visualization,polak2022efficiency}. Among these, the number of molecules with the FCC structure is negligible. We therefore conclude that D$_2$ undergoes single-crystal growth on the substrate with a lattice constant of 3.5 Å, whereas the grown structures on substrates with other lattice constants exhibit a distinct polycrystalline feature.

\begin{figure*}
	\centering
	\includegraphics[width=1.0\textwidth]{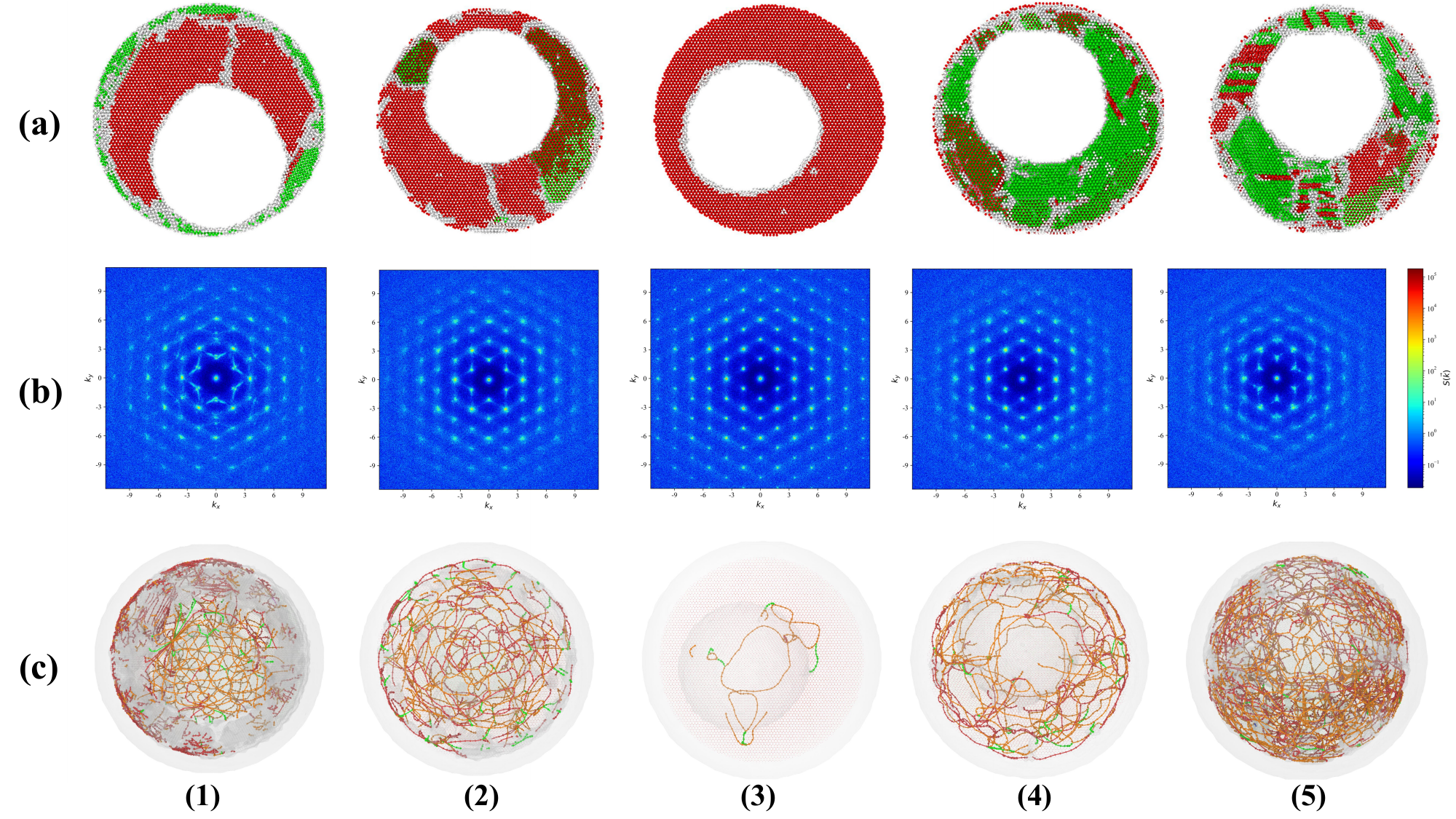} 
	\caption{(a) Schematic diagram of D$_2$ structure after the completion of crystallization under different substrates. Different colors represent distinct phase structures, where red denotes the HCP phase, green denotes the FCC phase, and white denotes the ``other" phase. Viewed along the $z$-axis with the substrate removed. (b) Diffraction patterns (2D static structure factors) of D$_2$ after the completion of crystallization under different substrates. (c) Dislocation lines of D$_2$ after the completion of crystallization under different substrates. Note: (1), (2), (3), (4), and (5) correspond to substrates with lattice constants of 3.1 Å, 3.3 Å, 3.5 Å, 3.7 Å, and 3.9 Å, respectively.} 
	\label{final}
\end{figure*}

Fig.~\ref{final} (a) presents the schematic diagram of the D$_2$ molecular structure after the completion of the crystallization process for substrates with different lattice constants, where different phases are marked with distinct colors. Among them, red denotes the HCP phase; green denotes the FCC phase; and white denotes the phase identified as ``other'' by PTM. It can be observed that D$_2$ undergoes single-crystal growth only when the substrate lattice constant is 3.5 Å; for other substrates, a large number of other distinct phases are observed.

Meanwhile, to further verify our conclusion, we computed the diffraction pattern—a 2D representation of the static structure factor $S(\vec{k})$ for this atomic system—which was performed using the freud package \cite{ramasubramani2020freud}. The results are shown in Fig.~\ref{final} (b). It can be observed that only when the substrate lattice constant is 3.5 Å does the diffraction pattern of D$_2$ exhibit a relatively strict six-fold symmetry (with diffraction spots arranged in a regular hexagonal pattern), a characteristic corresponding to the hexagonal reciprocal lattice; whereas for substrates with other lattice constants, the diffraction pattern of D$_2$ lacks a strict six-fold symmetry in comparison to that at 3.5 Å, indicating the occurrence of polycrystalline growth in the system. Furthermore, as the lattice mismatch degree increases, the characteristic of regular hexagonal distribution in the diffraction pattern becomes increasingly indistinct.

In addition, we also identified the dislocation characteristics of D$_2$ grown on different substrates using the OVITO software package \cite{stukowski2009visualization}, with the dislocation structures presented in Fig.~\ref{final} (c). It can be observed that when the lattice constant of the substrate is 3.5 Å, the number of dislocation lines is minimized. As the lattice mismatch degree increases, the number of dislocation lines rises significantly, indicating that a large number of dislocations are present in the finally grown D$_2$ ice structure. The dislocation density data are available in the supporting information.

It is also found that the D$_2$ grown on the substrate with a lattice constant of 3.5 Å still exhibits a certain number of dislocations. This is because the curvature effect hinders the ideal planar spreading of crystals during growth, inducing slight lattice distortion and thus inevitably introducing a small number of dislocations, which results in the presence of a small number of dislocation lines in the D$_2$ ice on this substrate.

To further investigate the crystallization quality of D$_2$ inside spherical shells with different lattice parameters, we first extracted the inner surface molecules of D$_2$ ice, then analyzed the power spectral density (PSD) of the distances from these inner surface molecules to the center of the sphere, and further derived the inner surface roughness. To ensure the accuracy and standardization of the PSD calculation, the specific procedure is carried out as follows.

First, inner surface atoms are identified using OVITO software \cite{stukowski2009visualization} by computing the coordination number (CN); molecules with a CN less than 11 are defined as inner surface atoms. Second, the center of the inner surface is determined by fitting the extracted molecules using the least squares method. Then, a latitude-longitude angular grid is constructed to select the molecule within each grid cell that is closest to the fitted center; these selected molecules constitute the final set used for subsequent analysis.

The power spectral density is calculated based on the spherical harmonic expansion method, with the detailed process referencing Ref.~\cite{lei2015solidification}. We established a system for characterizing the spatial positions of molecules in the spherical coordinate system: with the geometric center of the spherical shell as the origin, we defined the polar angle of any D$_2$ molecule on the inner surface as $\theta$, the azimuthal angle as $\phi$, and its distance from the origin as $R(\theta,\phi)$. The calculation of the power spectral density was implemented based on the spherical harmonic expansion method: specifically, $R(\theta,\phi)$ was first decomposed into a linear combination of a series of orthonormal spherical harmonic functions $Y_{lm}(\theta,\phi)$ \cite{lei2015solidification,kozioziemski2011deuterium}, and the expression is given by:
\begin{equation}
	R(\theta,\phi)=\sum_{l=0}^\infty\sum_{m=-l}^lA_{lm}Y_{lm}(\theta,\phi),
\end{equation}
this yields the power spectrum of the inner surface of the D$_2$ solid phase with mode number $l$:
\begin{equation}
	P_l=\frac{1}{4\pi}\sum_{m=-l}^l|A_{lm}|^2,
\end{equation}
the root-mean-square (rms) roughness $\sigma_{rms}$ of the inner surface of the D$_2$ solid phase can be computed in accordance with the following formula:
\begin{equation}
	\sigma_{\mathrm{rms}}=\sqrt{\sum_{l=1}^{l_{\mathrm{max}}}P_l}.\label{sigmarms}
\end{equation}
considering the characteristics of the inner surface of D$_2$ ice, and consistent with Ref.~\cite{lei2015solidification}, this study selects $l_{\mathrm{max}}=128$, which can fully cover the fluctuation information ranging from macroscopic shape to microscopic roughness.

\begin{figure}
	\centering
	\includegraphics[width=0.48\textwidth]{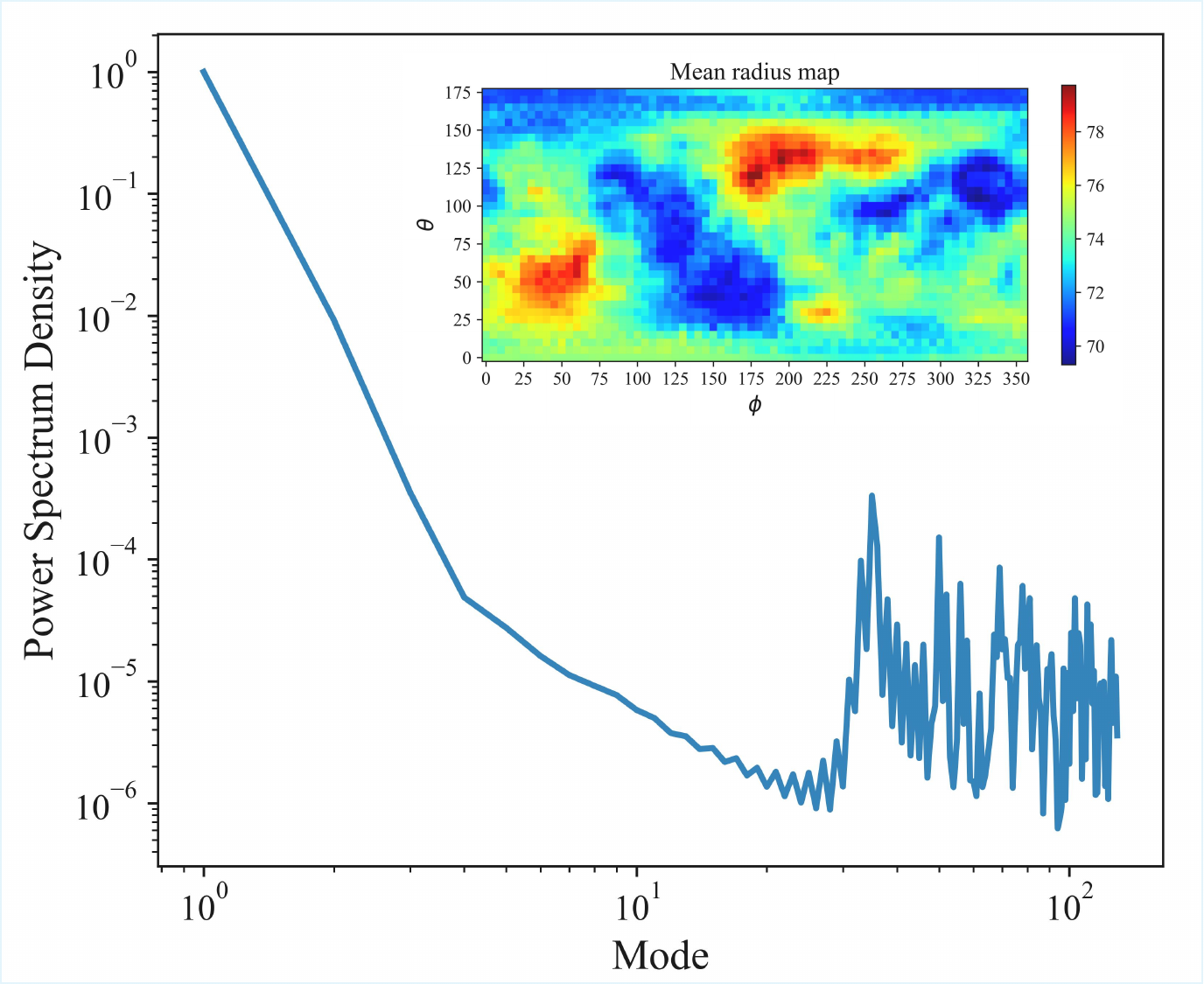} 
	\caption{Distance contour map of the inner surface molecules of the D$_2$ solid to the origin and the inner surface power spectrum for a substrate with a lattice constant of 3.5 Å.} 
	\label{psd}
\end{figure}

\begin{figure}
	\centering
	\includegraphics[width=0.48\textwidth]{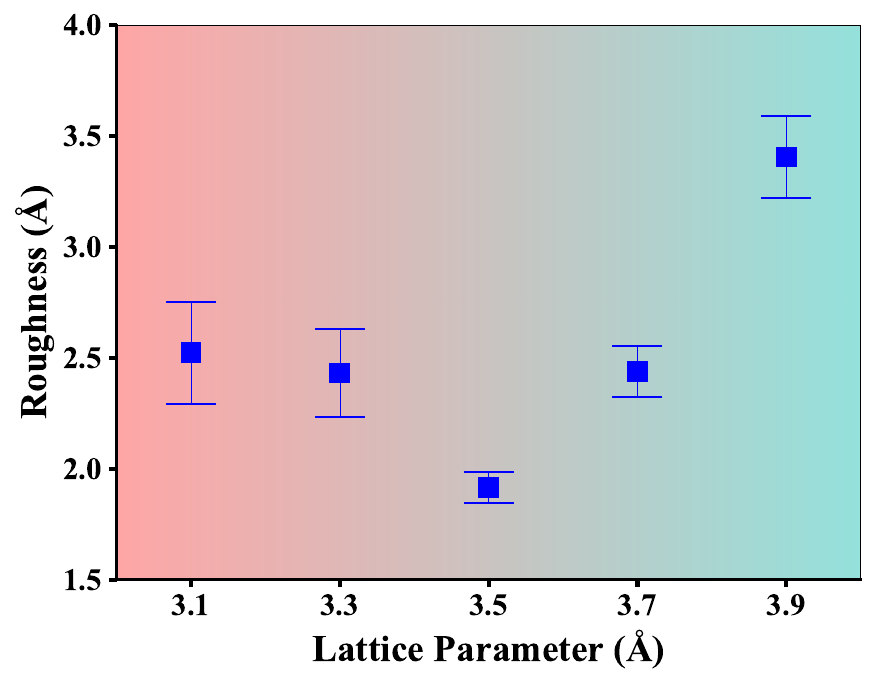} 
	\caption{Average inner surface roughness of solid D$_2$ grown on different substrates, obtained from five independent simulations.} 
	\label{rms}
\end{figure}

For a substrate with a lattice constant of 3.5 Å, the calculated distance contour map of the inner surface molecules of the D$_2$ solid to the origin and the inner surface power spectrum are shown in Fig.~\ref{psd}. Subsequently, we calculated the inner surface roughness of the D$_2$ solid under different substrates based on Eq.~\ref{sigmarms} and the results are presented in Fig.~\ref{rms}.

It is not difficult to observe from the results that the inner surface roughness increases significantly with the increase of the lattice mismatch degree, and the inner surface roughness is minimized when the substrate lattice constant is 3.5 Å. This phenomenon can be attributed to the regulatory effect of the lattice mismatch degree on the orderliness of molecular arrangement in the D$_2$ solid: when the lattice mismatch degree between the substrate and D$_2$ is relatively large, molecules tend to generate dislocations and accumulation due to lattice stress during the growth process. This leads to the formation of significant undulations on the inner surface, thereby increasing the roughness; additionally, this also affects the final result of crystallization growth, resulting in the formation of polycrystals. In contrast, when the substrate lattice constant is 3.5 Å, the lattice mismatch degree reaches the minimum. Under this condition, D$_2$ molecules can achieve a more orderly arrangement along the substrate surface, the molecular distribution on the inner surface is more uniform, and it is easier to realize single-crystal growth.

In addition, simulations show that after complete crystallization, the solid D$_2$ ice layer is slightly eccentric relative to the spherical shell substrate, and the spatial orientation of this eccentricity exhibits obvious randomness in multiple independent simulations; this randomness indicates that the initial nucleation process of D$_2$ on the substrate surface possesses an intrinsic stochastic characteristic, while the subsequent crystal growth proceeds isotropically, and such random nucleation behavior is mainly ascribed to local fluctuations induced by the random distribution of particle velocities in molecular dynamics simulations, which results in a probabilistic distribution of nucleation rates at different positions on the spherical substrate interface; although the statistical distribution of nucleation sites gives rise to minor eccentricity in the macroscopic structure, the key conclusions concerning crystallization quality and growth mechanisms remain highly consistent across all parallel simulations, and such secondary structural deviations exert no substantial impact on the core findings of the present study regarding the crystallization dynamics of D$_2$.

Meanwhile, the present atomic-scale simulations are performed under ideal isothermal and gravity-free conditions, which inherently eliminate the interference of external macroscopic factors such as thermal gradients and gravitational fields. Consequently, the slight eccentricity observed arises purely from the intrinsic dynamic randomness inherent in the solidification process of the quantum fluid.

Based on a comprehensive analysis of crystallization quality, it is evident that the lattice mismatch between the substrate and D$_2$ plays a key role in determining the final structural integrity and surface smoothness of the solid ice layer. When the substrate lattice constant is 3.5 Å, the system exhibits minimal lattice mismatch, leading to a highly ordered HCP single-crystalline structure characterized by very few dislocations and the lowest inner surface roughness.

In contrast, as the lattice mismatch increases, the system displays polycrystalline growth features with the coexistence of HCP and FCC phases: disordered molecular arrangements and the formation of stress-induced defects result in not only elevated dislocation density but also a significant increase in surface roughness.

\section{Summary and conclusion}
This study investigates the crystal growth mechanism of D$_2$ inside spherical shells with varying lattice constants, using MD simulations employing a quantum-corrected SG potential to account for NQE at low temperatures. The research aims to elucidate how substrate lattice matching influences the crystallization pathway, structural quality, and surface smoothness of the solid D$_2$ layer—a system analogous to deuterium–tritium (DT) fuel used in ICF.

The simulations suggest that lattice mismatch between the substrate and D$_2$ crystal significantly determines the growth mode and final quality of the ice layer. When the substrate lattice constant is 3.5 Å, closely matching the equilibrium lattice constant of solid D$_2$, the system exhibits a layer-by-layer growth mechanism consistent with Ostwald’s stepwise nucleation theory. This mode promotes the formation of a well-ordered HCP single crystal with minimal dislocations and the lowest inner surface roughness. This behavior arises from the lattice mismatch being minimized, which reduces the interfacial misfit strain, allows the curvature effect to largely govern the growth process, and thus enables layer-by-layer coherent epitaxial growth and the formation of low-defect single crystals.

In contrast, substrates with larger lattice mismatches (e.g., 3.1 Å, 3.3 Å, 3.7 Å, or 3.9 Å) induce an island-like growth mode, leading to polycrystalline structures with mixed HCP and FCC phases, high dislocation densities, and significantly increased surface roughness. This transition is suggested to be attributed to the significant lattice mismatch introducing compressive or tensile stress at the interface, which triggers plastic relaxation processes such as dislocations and stacking faults, introducing geometric frustration effects, disrupting epitaxial coherence, and thus transforming the growth mode into the island-like mode. The growth mode transition is accompanied by changes in the evolution of precursor and geometrically frustrated particles, with the fastest phase transition occurring when precursor concentration peaks. Meanwhile, further investigations via Polyhedral Template Matching, diffraction pattern analysis, and dislocation identification techniques illustrate that the substrate with optimal lattice matching (3.5 Å) enables the formation of a single-crystalline structure in D$_2$ crystals, whereas substrates with lattice mismatch induce the formation of polycrystalline structures and a significant degradation in their structural integrity.

Analysis of radial potential energy and stress distribution further clarifies the regulatory mechanism of lattice mismatch on crystal growth. Studies have shown that the interface stress induced by lattice mismatch is characterized by a significant localization and short-range effects, which rapidly decays to a negligible level within a range of approximately 2–3 molecular layers from the spherical shell interface. The interface stress is not a long-range dominant field governing the entire crystal growth process; instead, its main effect is reflected in the initial molecular layers near the interface: within this region, the interface stress determines the type of stress (tensile stress or compressive stress) and rapidly induces microstructural defects. Outside this thin interface region, external stress has been largely relaxed, and the subsequent crystal growth is mainly dominated by defects generated at the interface, such as misfit dislocations, grain boundaries, and clusters. This conclusion aligns well with the earlier observed island and layer-by-layer growth modes, further validating the regulatory role of lattice mismatch on growth mode and structural quality.

We also quantitatively characterized the inner surface roughness of solid D$_2$ via spherical harmonic expansion of the surface power spectral density, finding that it increases monotonically with increasing lattice mismatch, a result that highlights the critical role of substrate design in fabricating smooth and uniform solid fuel layers required for efficient ICF ignition.

In the present simulations, D$_2$ molecules are treated as spherically symmetric free rotors, a reasonable approximation at cryogenic temperatures. The spherical shell substrates are modeled as rigid lattices, serving solely as structural templates without accounting for atomic relaxation or interfacial chemical effects. While this work reveals a fundamental lattice‑matching mechanism within this simplified framework, extrapolation to more realistic systems such as DT mixtures, compliant shells, or chemically active interfaces requires further validation. Future studies incorporating substrate compliance and multi‑component fuels will be necessary to assess the full relevance of the proposed mechanism for ICF applications.

Although our model system adopts idealised spherical geometry and a fixed cooling rate, representing simplifications relative to real ICF target environments involving complex thermomechanical histories and larger scales, the identified mechanism of lattice-match-controlled growth provides clear guidance for substrate engineering in cryogenic target fabrication, highlighting that precise substrate lattice tuning is a key factor for growing high-quality, single-crystal D$_2$ ice layers with minimal defects and optimal surface smoothness. These findings provide valuable insights for the design of cryogenic targets in ICF, particularly for the preparation of high-quality DT fuel layers \cite{sater2016technique,walters2018d2}.

\begin{acknowledgments}
The authors thank Professor Hai-Le Lei for the useful discussions.
\end{acknowledgments}

\section*{Author Declarations}
\subsection*{Conflict of Interest}
The authors have no conflicts to disclose.

\subsection*{Author Contributions}
Yu-Shen Wan and Peng Bi contributed equally.

{\bf Peng Bi:} Conceptualization (lead); Formal Analysis (equal); Investigation (lead); Visualization (lead); Writing – original draft (supporting); Writing – review \& editing (lead); Project administration (lead). {\bf Yu-Shen Wan:} Formal Analysis (equal); Software (lead); Investigation (equal); Validation (lead); Writing – original draft (lead); Writing – review \& editing (equal). {\bf Wei Zhang:} Investigation (supporting); Data curation (lead); Formal Analysis (supporting); Validation (supporting). {\bf Jian Chen:} Investigation (supporting); Software (supporting); Validation (supporting); Visualization (supporting). {\bf Yong Yi:} Conceptualization (supporting); Resources (lead); Supervision (lead); Project administration (supporting); Writing – review \& editing (equal). {\bf Qi-Feng Chen:} Conceptualization (supporting); Resources (supporting); Supervision (lead); Writing – review \& editing (lead).

\section*{Data Availability Statement}
The data that support the findings of this study are available from the corresponding author upon reasonable request.

%\section*{References}
\bibliography{newref}

@article{kozioziemski2010metastable,
	title={Metastable polymorphs of hydrogen isotopes solidified near the triple point},
	author={Kozioziemski, BJ and Chernov, AA and Mapoles, ER and Sater, JD},
	journal={Physical Review B—Condensed Matter and Materials Physics},
	volume={82},
	number={1},
	pages={012104},
	year={2010},
	publisher={APS}
}

@article{markland2018nuclear,
	title={Nuclear quantum effects enter the mainstream},
	author={Markland, Thomas E and Ceriotti, Michele},
	journal={Nature Reviews Chemistry},
	volume={2},
	number={3},
	pages={0109},
	year={2018},
	publisher={Nature Publishing Group UK London}
}

@article{shin2016supercooling,
	title={Supercooling of hydrogen on template materials to deterministically seed ignition-quality solid fuel layers},
	author={Shin, SJ and Zepeda-Ruiz, LA and Lee, JRI and Baxamusa, SH and Dylla-Spears, R and Suratwala, T and Kozioziemski, BJ},
	journal={Fusion Science and Technology},
	volume={70},
	number={2},
	pages={184--190},
	year={2016},
	publisher={Taylor \& Francis}
}

@article{lei2015solidification,
	title={Solidification of an atomic fluid inside a spherical shell},
	author={Lei, Haile and Bi, Peng and Yi, Yong and Wang, Kai and Lin, Wei},
	journal={Nuclear Fusion},
	volume={55},
	number={6},
	pages={063033},
	year={2015},
	publisher={IOP Publishing}
}

@article{zepeda2018effect,
	title={Effect of wetting on nucleation and growth of D2 in confinement},
	author={Zepeda-Ruiz, LA and Sadigh, B and Shin, SJ and Kozioziemski, BJ and Chernov, AA},
	journal={The Journal of Chemical Physics},
	volume={148},
	number={13},
	year={2018},
	publisher={AIP Publishing}
}

@article{shin2018materials,
	title={Materials and Morphology Study for Templated Hydrogen Solidification},
	author={Shin, Swanee J and Kozioziemski, Bernard J},
	journal={Fusion Science and Technology},
	volume={73},
	number={3},
	pages={298--304},
	year={2018},
	publisher={Taylor \& Francis}
}

@article{kozioziemski1997crystal,
	title={Crystal growth and roughening of solid D2},
	author={Kozioziemski, BJ and Collins, GW and Bernat, TP},
	journal={Fusion Technology},
	volume={31},
	number={4},
	pages={482--484},
	year={1997},
	publisher={Taylor \& Francis}
}

@article{silvera1978isotropic,
	title={The isotropic intermolecular potential for H2 and D2 in the solid and gas phases},
	author={Silvera, Isaac F and Goldman, Victor V},
	journal={The Journal of Chemical Physics},
	volume={69},
	number={9},
	pages={4209--4213},
	year={1978},
	publisher={American Institute of Physics}
}

@article{aasen2019equation,
	title={Equation of state and force fields for Feynman--Hibbs-corrected Mie fluids. I. Application to pure helium, neon, hydrogen, and deuterium},
	author={Aasen, Ailo and Hammer, Morten and Ervik, {\AA}smund and M{\"u}ller, Erich A and Wilhelmsen, {\O}ivind},
	journal={The Journal of Chemical Physics},
	volume={151},
	number={6},
	year={2019},
	publisher={AIP Publishing}
}

@book{feynman2005quantum,
	title={Quantum mechanics and path integrals: Emended edition},
	author={Feynman, Richard Phillips and Hibbs, Albert R and Styer, Daniel F},
	year={2005},
	publisher={Dover Publications}
}

@article{jervell2025limits,
	title={The limits of Feynman--Hibbs corrections in capturing quantum-nuclear contributions to thermophysical properties},
	author={Jervell, Vegard G and Wilhelmsen, {\O}ivind},
	journal={The Journal of Chemical Physics},
	volume={163},
	number={14},
	year={2025},
	publisher={AIP Publishing}
}

@article{lenhard2024history,
	title={On the History of the Lennard-Jones Potential},
	author={Lenhard, Johannes and Stephan, Simon and Hasse, Hans},
	journal={Annalen der Physik},
	volume={536},
	number={6},
	pages={2400115},
	year={2024},
	publisher={Wiley Online Library}
}

@article{thompson2022lammps,
	title={LAMMPS-a flexible simulation tool for particle-based materials modeling at the atomic, meso, and continuum scales},
	author={Thompson, Aidan P and Aktulga, H Metin and Berger, Richard and Bolintineanu, Dan S and Brown, W Michael and Crozier, Paul S and In't Veld, Pieter J and Kohlmeyer, Axel and Moore, Stan G and Nguyen, Trung Dac and others},
	journal={Computer physics communications},
	volume={271},
	pages={108171},
	year={2022},
	publisher={Elsevier}
}

@article{zhan2023multiple,
	title={Multiple scenarios of low-temperature nucleation in simple liquids},
	author={Zhan, Mengyuan and Chen, Yanshuang and Jiang, Zhehua and Xu, Ning and Tan, Peng},
	journal={Physical Review Letters},
	volume={130},
	number={17},
	pages={178201},
	year={2023},
	publisher={APS}
}

@article{steinhardt1983bond,
	title={Bond-orientational order in liquids and glasses},
	author={Steinhardt, Paul J and Nelson, David R and Ronchetti, Marco},
	journal={Physical Review B},
	volume={28},
	number={2},
	pages={784},
	year={1983},
	publisher={APS}
}

@article{tan2014visualizing,
	title={Visualizing kinetic pathways of homogeneous nucleation in colloidal crystallization},
	author={Tan, Peng and Xu, Ning and Xu, Lei},
	journal={Nature Physics},
	volume={10},
	number={1},
	pages={73--79},
	year={2014},
	publisher={Nature Publishing Group UK London}
}

@article{rein1996numerical,
	title={Numerical calculation of the rate of crystal nucleation in a Lennard-Jones system at moderate undercooling},
	author={Rein ten Wolde, Pieter and Ruiz-Montero, Maria J and Frenkel, Daan},
	journal={The Journal of chemical physics},
	volume={104},
	number={24},
	pages={9932--9947},
	year={1996},
	publisher={American Institute of Physics}
}

@article{lechner2008accurate,
	title={Accurate determination of crystal structures based on averaged local bond order parameters},
	author={Lechner, Wolfgang and Dellago, Christoph},
	journal={The Journal of chemical physics},
	volume={129},
	number={11},
	year={2008},
	publisher={AIP Publishing}
}

@article{ten1999homogeneous,
	title={Homogeneous nucleation and the Ostwald step rule},
	author={Ten Wolde, Pieter Rein and Frenkel, Daan},
	journal={Physical Chemistry Chemical Physics},
	volume={1},
	number={9},
	pages={2191--2196},
	year={1999},
	publisher={Royal Society of Chemistry}
}

@incollection{schmelzer2017crystals,
	title={How do crystals nucleate and grow: Ostwald’s rule of stages and beyond},
	author={Schmelzer, J{\"u}rn WP and Abyzov, Alexander S},
	booktitle={Thermal physics and thermal analysis: From macro to micro, highlighting thermodynamics, kinetics and nanomaterials},
	pages={195--211},
	year={2017},
	publisher={Springer}
}

@article{lozovoy2020kinetics,
	title={Kinetics of epitaxial formation of nanostructures by Frank--van der Merwe, Volmer--Weber and Stranski--Krastanow growth modes},
	author={Lozovoy, Kirill A and Korotaev, Alexander G and Kokhanenko, Andrey P and Dirko, Vladimir V and Voitsekhovskii, Alexander V},
	journal={Surface and Coatings Technology},
	volume={384},
	pages={125289},
	year={2020},
	publisher={Elsevier}
}

@article{velikov2002layer,
	title={Layer-by-layer growth of binary colloidal crystals},
	author={Velikov, Krassimir P and Christova, Christina G and Dullens, Roel PA and van Blaaderen, Alfons},
	journal={Science},
	volume={296},
	number={5565},
	pages={106--109},
	year={2002},
	publisher={American Association for the Advancement of Science}
}

@article{puurunen2004island,
	title={Island growth as a growth mode in atomic layer deposition: A phenomenological model},
	author={Puurunen, Riikka L and Vandervorst, Wilfried},
	journal={Journal of Applied Physics},
	volume={96},
	number={12},
	pages={7686--7695},
	year={2004},
	publisher={AIP Publishing}
}

@article{polak2022efficiency,
	title={Efficiency in identification of internal structure in simulated monoatomic clusters: Comparison between common neighbor analysis and coordination polyhedron method},
	author={Polak, Wies{\l}aw Z},
	journal={Computational Materials Science},
	volume={201},
	pages={110882},
	year={2022},
	publisher={Elsevier}
}

@article{ramasubramani2020freud,
	title={freud: A software suite for high throughput analysis of particle simulation data},
	author={Ramasubramani, Vyas and Dice, Bradley D and Harper, Eric S and Spellings, Matthew P and Anderson, Joshua A and Glotzer, Sharon C},
	journal={Computer Physics Communications},
	volume={254},
	pages={107275},
	year={2020},
	publisher={Elsevier}
}

@article{stukowski2009visualization,
	title={Visualization and analysis of atomistic simulation data with OVITO--the Open Visualization Tool},
	author={Stukowski, Alexander},
	journal={Modelling and simulation in materials science and engineering},
	volume={18},
	number={1},
	pages={015012},
	year={2009},
	publisher={IOP Publishing}
}

@article{kozioziemski2011deuterium,
	title={Deuterium-tritium fuel layer formation for the National Ignition Facility},
	author={Kozioziemski, BJ and Mapoles, ER and Sater, JD and Chernov, AA and Moody, JD and Lugten, JB and Johnson, MA},
	journal={Fusion Science and Technology},
	volume={59},
	number={1},
	pages={14--25},
	year={2011},
	publisher={Taylor \& Francis}
}

@article{sater2016technique,
	title={Technique for Forming Solid D2 and DT Layers for Shock Timing Experiments at the National Ignition Facility},
	author={Sater, JD and Espinosa-Loza, F and Kozioziemski, B and Mapoles, ER and Dylla-Spears, R and Pipes, JW and Walters, CF},
	journal={Fusion Science and Technology},
	volume={70},
	number={2},
	pages={191--195},
	year={2016},
	publisher={Taylor \& Francis}
}

@article{walters2018d2,
	title={D2 and dT liquid-layer target shots at the national ignition facility},
	author={Walters, Curtis and Alger, Ethan and Bhandarkar, Suhas and Boehm, Kurt and Braun, Tom and Espinosaloza, Francisco and Haid, Benjamin and Heredia, Ricardo and Kline, John and Kozioziemski, Bernard and others},
	journal={Fusion Science and Technology},
	volume={73},
	number={3},
	pages={305--314},
	year={2018},
	publisher={Taylor \& Francis}
}

@article{mickel2013shortcomings,
	title={Shortcomings of the bond orientational order parameters for the analysis of disordered particulate matter},
	author={Mickel, Walter and Kapfer, Sebastian C and Schr{\"o}der-Turk, Gerd E and Mecke, Klaus},
	journal={The Journal of chemical physics},
	volume={138},
	number={4},
	year={2013},
	publisher={AIP Publishing}
}

@article{arai2017surface,
	title={Surface-assisted single-crystal formation of charged colloids},
	author={Arai, Shunto and Tanaka, Hajime},
	journal={Nature Physics},
	volume={13},
	number={5},
	pages={503--509},
	year={2017},
	publisher={Nature Publishing Group UK London}
}

@article{larsen2016robust,
	title={Robust structural identification via polyhedral template matching},
	author={Larsen, Peter Mahler and Schmidt, S{\o}ren and Schi{\o}tz, Jakob},
	journal={Modelling and Simulation in Materials Science and Engineering},
	volume={24},
	number={5},
	pages={055007},
	year={2016},
	publisher={IOP Publishing}
}

@article{zhou2003new,
	title={A new look at the atomic level virial stress: on continuum-molecular system equivalence},
	author={Zhou, Min},
	journal={Proceedings of the Royal Society of London. Series A: Mathematical, Physical and Engineering Sciences},
	volume={459},
	number={2037},
	pages={2347--2392},
	year={2003},
	publisher={The Royal Society}
}

@article{chen2016origin,
	title={The origin of the distinction between microscopic formulas for stress and Cauchy stress},
	author={Chen, Youping},
	journal={Europhysics Letters},
	volume={116},
	number={3},
	pages={34003},
	year={2016},
	publisher={IOP Publishing}
}

@article{subramaniyan2008continuum,
	title={Continuum interpretation of virial stress in molecular simulations},
	author={Subramaniyan, Arun K and Sun, CT},
	journal={International Journal of Solids and Structures},
	volume={45},
	number={14-15},
	pages={4340--4346},
	year={2008},
	publisher={Elsevier}
}

@article{xu2009investigation,
	title={Investigation on applicability of various stress definitions in atomistic simulation},
	author={Xu, Ran and Liu, Bin},
	journal={Acta Mechanica Solida Sinica},
	volume={22},
	number={6},
	pages={644--649},
	year={2009},
	publisher={Springer}
}

@article{liu2009compute,
	title={How to compute the atomic stress objectively?},
	author={Liu, Bin and Qiu, Xinming},
	journal={Journal of Computational and Theoretical Nanoscience},
	volume={6},
	number={5},
	pages={1081--1089},
	year={2009},
	publisher={American Scientific Publishers}
}

@article{dai2021comparative,
	title={A comparative study of atomistic-based stress evaluation},
	author={Dai, Shuyang and Wang, Fengru and Yang, Jerry Zhijian and Yuan, Cheng},
	journal={Discrete and Continuous Dynamical Systems-B},
	volume={26},
	number={9},
	pages={4999--5021},
	year={2021},
	publisher={Discrete and Continuous Dynamical Systems-B}
}

@article{morris1994melting,
	title={Melting line of aluminum from simulations of coexisting phases},
	author={Morris, James R and Wang, CZ and Ho, KM and Chan, Che Ting},
	journal={Physical Review B},
	volume={49},
	number={5},
	pages={3109},
	year={1994},
	publisher={APS}
}

@article{ogitsu2003melting,
	title={Melting of lithium hydride under pressure},
	author={Ogitsu, Tadashi and Schwegler, Eric and Gygi, Francois and Galli, Giulia},
	journal={Physical review letters},
	volume={91},
	number={17},
	pages={175502},
	year={2003},
	publisher={APS}
}

@article{schwalbe1984pressure,
	title={Pressure-volume-temperature relationships for normal deuterium between 18.7 and 21.0 K},
	author={Schwalbe, LA and Grilly, ER},
	journal={JOURNAL OF RESEARCH of the National Bureau of Standards},
	volume={89},
	number={3},
	pages={227},
	year={1984}
}

@article{ali2018hydrodynamic,
  title={Hydrodynamic instability seeding by oxygen nonuniformities in glow discharge polymer inertial fusion ablators},
  author={Ali, SJ and Celliers, PM and Haan, SW and Boehly, TR and Whiting, N and Baxamusa, SH and Reynolds, H and Johnson, MA and Hughes, JD and Watson, B and others},
  journal={Physical Review E},
  volume={98},
  number={3},
  pages={033204},
  year={2018},
  publisher={APS}
}

@article{du2018recent,
  title={Recent progress in ICF target fabrication at RCLF},
  author={Du, Kai and Liu, Meifang and Wang, Tao and He, Xiaoshan and Wang, Zongwei and Zhang, Juan},
  journal={Matter and Radiation at Extremes},
  volume={3},
  number={3},
  pages={135--144},
  year={2018},
  publisher={AIP Publishing}
}

@inproceedings{ho2016implosion,
  title={Implosion configurations for robust ignition using high-density carbon (diamond) ablator for indirect-drive ICF at the National Ignition Facility},
  author={Ho, DD-M and Haan, SW and Salmonson, JD and Clark, DS and Lindl, JD and Milovich, JL and Thomas, CA and Berzak Hopkins, LF and Meezan, NB},
  booktitle={Journal of Physics: Conference Series},
  volume={717},
  number={1},
  pages={012023},
  year={2016},
  organization={IOP Publishing}
}

@article{wang2021density,
  title={Density-dependent shock Hugoniot of polycrystalline diamond at pressures relevant to ICF},
  author={Wang, Peng and Zhang, Chen and Jiang, Shaoen and Duan, Xiaoxi and Zhang, Huan and Li, LiLing and Yang, Weiming and Liu, Yonggang and Li, Yulong and Sun, Liang and others},
  journal={Matter and Radiation at Extremes},
  volume={6},
  number={3},
  year={2021},
  publisher={AIP Publishing}
}

@article{kritcher2018comparison,
  title={Comparison of plastic, high density carbon, and beryllium as indirect drive NIF ablators},
  author={Kritcher, AL and Clark, D and Haan, S and Yi, SA and Zylstra, AB and Callahan, DA and Hinkel, DE and Berzak Hopkins, LF and Hurricane, OA and Landen, OL and others},
  journal={Physics of Plasmas},
  volume={25},
  number={5},
  year={2018},
  publisher={AIP Publishing}
}

@article{luo2017investigation,
  title={An investigation progress toward Be-based ablator materials for the inertial confinement fusion},
  author={Luo, Bingchi and Zhang, Jiqiang and He, Yudan and Chen, Long and Luo, Jiangshan and Li, Kai and Wu, Weidong},
  journal={High Power Laser Science and Engineering},
  volume={5},
  pages={e10},
  year={2017},
  publisher={Cambridge University Press}
}

@article{cao2020beryllium,
  title={Beryllium carbide as diffusion barrier against Cu: First-principles study},
  author={Cao, Hua-Liang and Cheng, Xin-Lu and Zhang, Hong},
  journal={Chinese Physics B},
  volume={29},
  number={1},
  pages={016601},
  year={2020},
  publisher={Chinese Physical Society and IOP Publishing Ltd}
}

@article{yan2025hydrodynamic,
  title={Hydrodynamic instability growth of the fuel--ablator interface induced by rippled rarefaction waves in inertial confinement fusion implosion experiments},
  author={Yan, Zheng and Chen, Zhu and Li, Jiwei and Wang, Lifeng and Li, Zhiyuan and Zhang, Chao and Ge, Fengjun and Wu, Junfeng and Zhang, Weiyan},
  journal={Matter and Radiation at Extremes},
  volume={10},
  number={5},
  year={2025},
  publisher={AIP Publishing}
}

@article{yang2026scaling,
  title={Scaling of thin wire cylindrical compression with material, diameter, and laser energy after 100 fs Joule surface heating},
  author={Yang, L and Herbert, M-L and Baehtz, C and Bouffetier, V and Brambrink, E and Dornheim, T and Fefeu, N and Gawne, T and Goede, S and Hagemann, J and others},
  journal={Matter and Radiation at Extremes},
  volume={11},
  number={1},
  year={2026},
  publisher={AIP Publishing}
}

@article{yang2021analyzing,
  title={Analyzing and relieving the thermal issues caused by fabrication details of a deuterium cryogenic target},
  author={Yang, Hong and Gao, Shasha and Jiang, Baibin and Xie, Jun and Liang, Juxi and Qi, Xiaobo and Wang, Kai and Tao, Chaoyou and Dai, Fei and Lin, Wei and others},
  journal={Matter and Radiation at Extremes},
  volume={6},
  number={5},
  year={2021},
  publisher={AIP Publishing}
}

@article{cipriani2026experimental,
  title={Experimental and simulation study on high-power laser irradiation of 3D-printed microstructures},
  author={Cipriani, M and Consoli, F and Scisci{\'o}, M and Solovjovas, A and Petsi, IA and Malinauskas, M and Andreoli, P and Cristofari, G and Di Ferdinando, E and Di Giorgio, G},
  journal={Matter and Radiation at Extremes},
  volume={11},
  number={2},
  year={2026},
  publisher={AIP Publishing}
}

@article{lindl2004physics,
  title={The physics basis for ignition using indirect-drive targets on the National Ignition Facility},
  author={Lindl, John D and Amendt, Peter and Berger, Richard L and Glendinning, S Gail and Glenzer, Siegfried H and Haan, Steven W and Kauffman, Robert L and Landen, Otto L and Suter, Laurence J},
  journal={Physics of plasmas},
  volume={11},
  number={2},
  pages={339--491},
  year={2004},
  publisher={American Institute of Physics}
}

@article{haan2004design,
  title={Design and simulations of indirect drive ignition targets for NIF},
  author={Haan, SW and Amendt, PA and Dittrich, TR and Hammel, BA and Hatchett, SP and Herrmann, MC and Hurricane, OA and Jones, OS and Lindl, JD and Marinak, MM and others},
  journal={Nuclear fusion},
  volume={44},
  number={12},
  pages={S171--S176},
  year={2004}
}

@article{chernov2009single,
  title={Single crystal growth and formation of defects in deuterium-tritium layers for inertial confinement nuclear fusion},
  author={Chernov, AA and Kozioziemski, BJ and Koch, JA and Atherton, LJ and Johnson, MA and Hamza, AV and Kucheyev, SO and Lugten, JB and Mapoles, EA and Moody, JD and others},
  journal={Applied Physics Letters},
  volume={94},
  number={6},
  year={2009},
  publisher={AIP Publishing}
}

@incollection{kalikmanov2012classical,
  title={Classical nucleation theory},
  author={Kalikmanov, Vitaly I},
  booktitle={Nucleation theory},
  pages={17--41},
  year={2012},
  publisher={Springer}
}

@article{caillabet2011change,
  title={Change in Inertial Confinement Fusion Implosions upon Using<? format?> an Ab Initio Multiphase DT Equation of State},
  author={Caillabet, Laurent and Canaud, Benoit and Salin, Gwena{\"e}l and Mazevet, St{\'e}phane and Loubeyre, Paul},
  journal={Physical Review Letters},
  volume={107},
  number={11},
  pages={115004},
  year={2011},
  publisher={APS}
}

@article{kang2020unified,
  title={Unified first-principles equations of state of deuterium-tritium mixtures in the global inertial confinement fusion region},
  author={Kang, Dongdong and Hou, Yong and Zeng, Qiyu and Dai, Jiayu},
  journal={Matter and Radiation at Extremes},
  volume={5},
  number={5},
  year={2020},
  publisher={AIP Publishing}
}

@article{stukowski2010extracting,
  title={Extracting dislocations and non-dislocation crystal defects from atomistic simulation data},
  author={Stukowski, Alexander and Albe, Karsten},
  journal={Modelling and Simulation in Materials Science and Engineering},
  volume={18},
  number={8},
  pages={085001},
  year={2010}
}

@article{stukowski2012automated,
  title={Automated identification and indexing of dislocations in crystal interfaces},
  author={Stukowski, Alexander and Bulatov, Vasily V and Arsenlis, Athanasios},
  journal={Modelling and Simulation in Materials Science and Engineering},
  volume={20},
  number={8},
  pages={085007},
  year={2012},
  publisher={IOP Publishing}
}

\end{document}

\title{Supporting Information for\\Lattice Matching Dictates the Growth Mode and Quality of Deuterium Crystallization in Confined Spherical Shells}
\date{\today}

\author{Peng Bi}
\thanks{Author to whom correspondence should be addressed: bipeng010@swust.edu.cn, yiyong@swust.edu.cn, chenqf01@gmail.com.}
\affiliation{School of Mathematics and Physics, Southwest University of Science and Technology, Mianyang 621010, China}

\author{Yu-Shen Wan}
\thanks{Yu-Shen Wan and Peng Bi contributed equally.} 
\affiliation{Beijing National Laboratory for Condensed Matter Physics, Institute of Physics, Chinese Academy of Sciences, Beijing 100190, China}
\affiliation{College of Materials Science and Opto-Electronic Technology, University of Chinese Academy of Sciences, Beijing 100049, China}

\author{Wei Zhang}
\affiliation{School of Mathematics and Physics, Southwest University of Science and Technology, Mianyang 621010, China}

\author{Jian Chen}
\affiliation{School of Mathematics and Physics, Southwest University of Science and Technology, Mianyang 621010, China}

\author{Yong Yi}
\thanks{Author to whom correspondence should be addressed: bipeng010@swust.edu.cn, yiyong@swust.edu.cn, chenqf01@gmail.com.}
\affiliation{School of Materials and Chemistry, Southwest University of Science and Technology, Mianyang 621010, China}

\author{Qi-Feng Chen}
\thanks{Author to whom correspondence should be addressed: bipeng010@swust.edu.cn, yiyong@swust.edu.cn, chenqf01@gmail.com.}
\affiliation{School of Mathematics and Physics, Southwest University of Science and Technology, Mianyang 621010, China}
\affiliation{National Key Laboratory for Shock Wave and Detonation Physics Research, Institute of Fluid Physics, China Academy of Engineering Physics, Mianyang 621900, China}

\maketitle

\section{Calculation of the Melting Point of D$_2$ at Zero Pressure}
This study uses the two-phase coexistence method in molecular dynamics simulations to determine the melting point of solid D$_2$. The simulations were performed using the LAMMPS software package \cite{thompson2022lammps}. The interatomic interactions were described by the Silvera-Goldman potential with first-order Feynman-Hibbs quantum corrections.

First, we constructed an initial D$_2$ two-phase coexistence system containing solid and liquid phases. The system was energy-minimized using the conjugate gradient algorithm, followed by long-time relaxation in the NPT ensemble to simulate the phase equilibrium process. Temperature and pressure were maintained using the Nosé-Hoover thermostat and the Anderson/Hoover pressure control method, respectively, with both the temperature coupling constant and pressure coupling constant set to 0.1 ps. The total simulation duration for the equilibration stage was 10 ns, with a time step of 1 fs.

The melting point was determined based on the stability of the two-phase interface at a specific temperature, which was achieved by monitoring the time evolution of the total potential energy of the system. When the temperature was below the melting point, the solid phase region grew and the potential energy decreased; when above the melting point, the liquid phase region grew and the potential energy increased; near the melting point, the two phases coexisted with the interface remaining stable, and the properties of each phase did not exhibit systematic drift over time.

\begin{figure}
	\centering
	\includegraphics[width=0.78\textwidth]{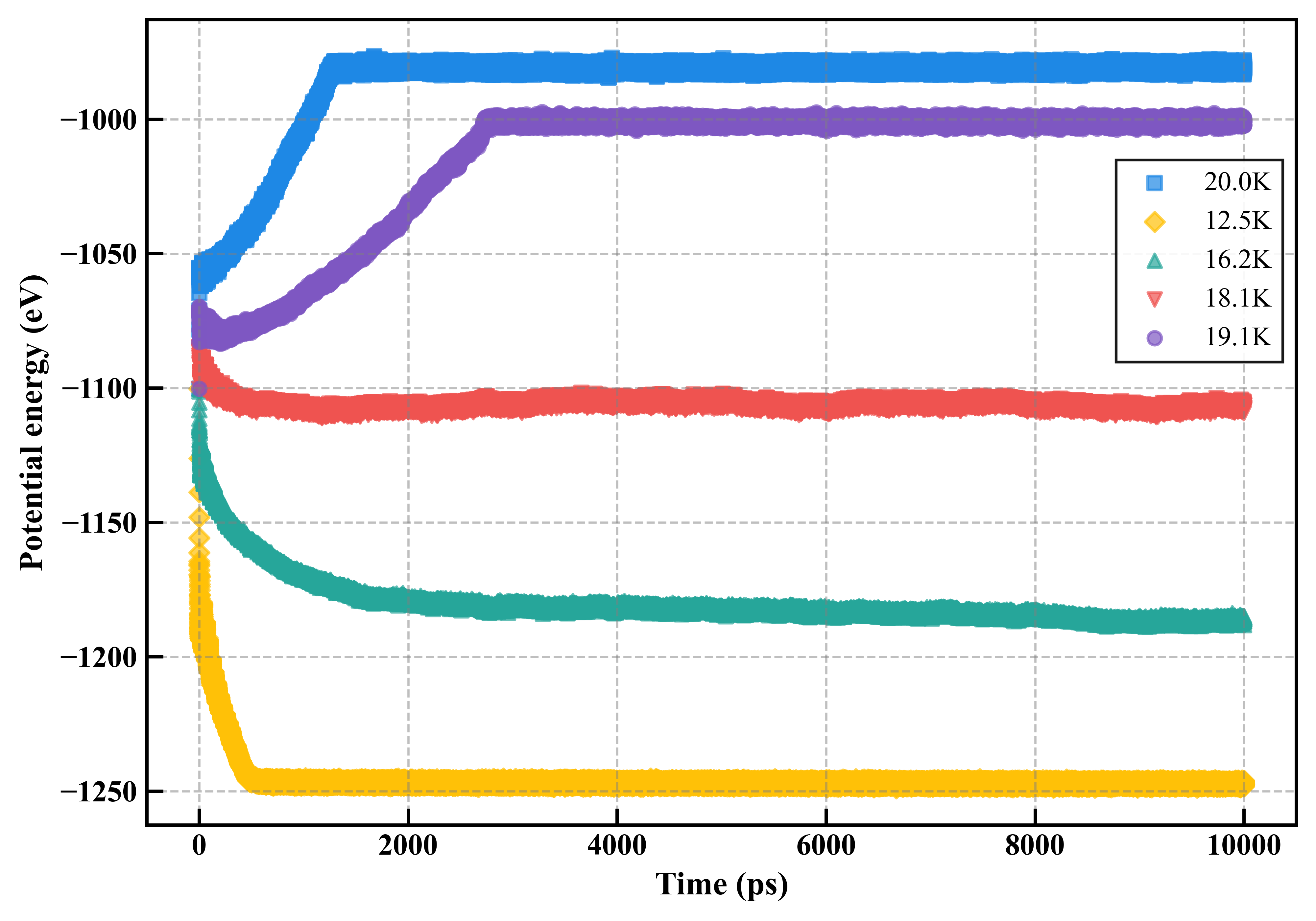} 
	\caption{Potential energy as a function of time for the D$_2$ solid-liquid coexistence system in a 12$\times$12$\times$120 supercell at different temperatures.} 
	\textbf{Alt text:} Potential energy versus time plot for a D$_2$ solid-liquid coexistence system in a 12$\times$12$\times$120 supercell at various temperatures. Only at 18.1 K does the potential energy exhibit neither a sharp increase (melting) nor a sharp decrease (crystallization).
	\label{pes}
\end{figure}

Fig.~\ref{pes} presents the potential energy time evolution of the D$_2$ solid-liquid coexistence system in a 12$\times$12$\times$120 supercell at various temperatures. The melting point of D$_2$ derived from the two-phase coexistence method is about 18.1 K.

\section{Equilibrium Lattice Constant of Solid D$_2$ at Different Temperatures}
\begin{figure}
	\centering
	\includegraphics[width=0.8\textwidth]{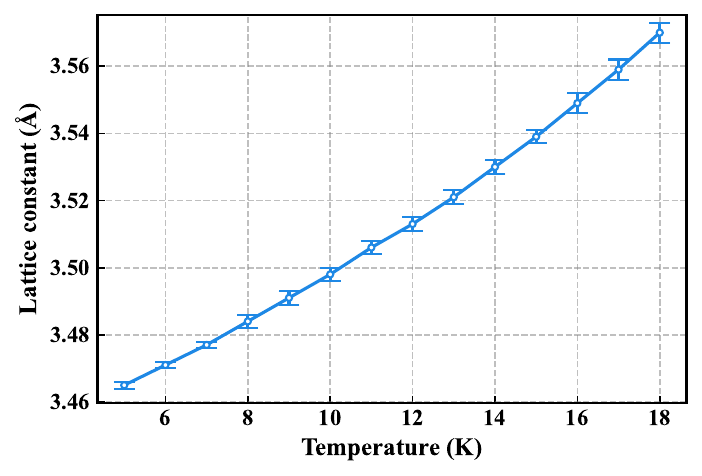} 
	\caption{Equilibrium lattice constant of solid D$_2$ as a function of temperature.}
	\textbf{Alt text:} Equilibrium lattice constant of solid D$_2$ plotted against temperature. It is 3.498 $\pm$ 0.002 Å at 10 K.
	\label{ltc}
\end{figure}

Based on the quantum-corrected Silvera-Goldman potential, we performed relaxation on the solid D$_2$ (HCP phase) under the NPT ensemble, and obtained the equilibrium lattice constants of solid D$_2$ at different temperatures, as shown in Fig.~\ref{ltc}.

\section{D$_2$ Crystal Growth Process}
As mentioned in the main text, the ratio of the proportion of precursor particles to that of geometrically frustrated particles is a very important parameter. Similar to D$_2$ grown on the 3.5 Å substrate, D$_2$ grown on all substrates exhibits a consistent trend.

\begin{figure}
	\centering
	\includegraphics[width=0.52\textwidth]{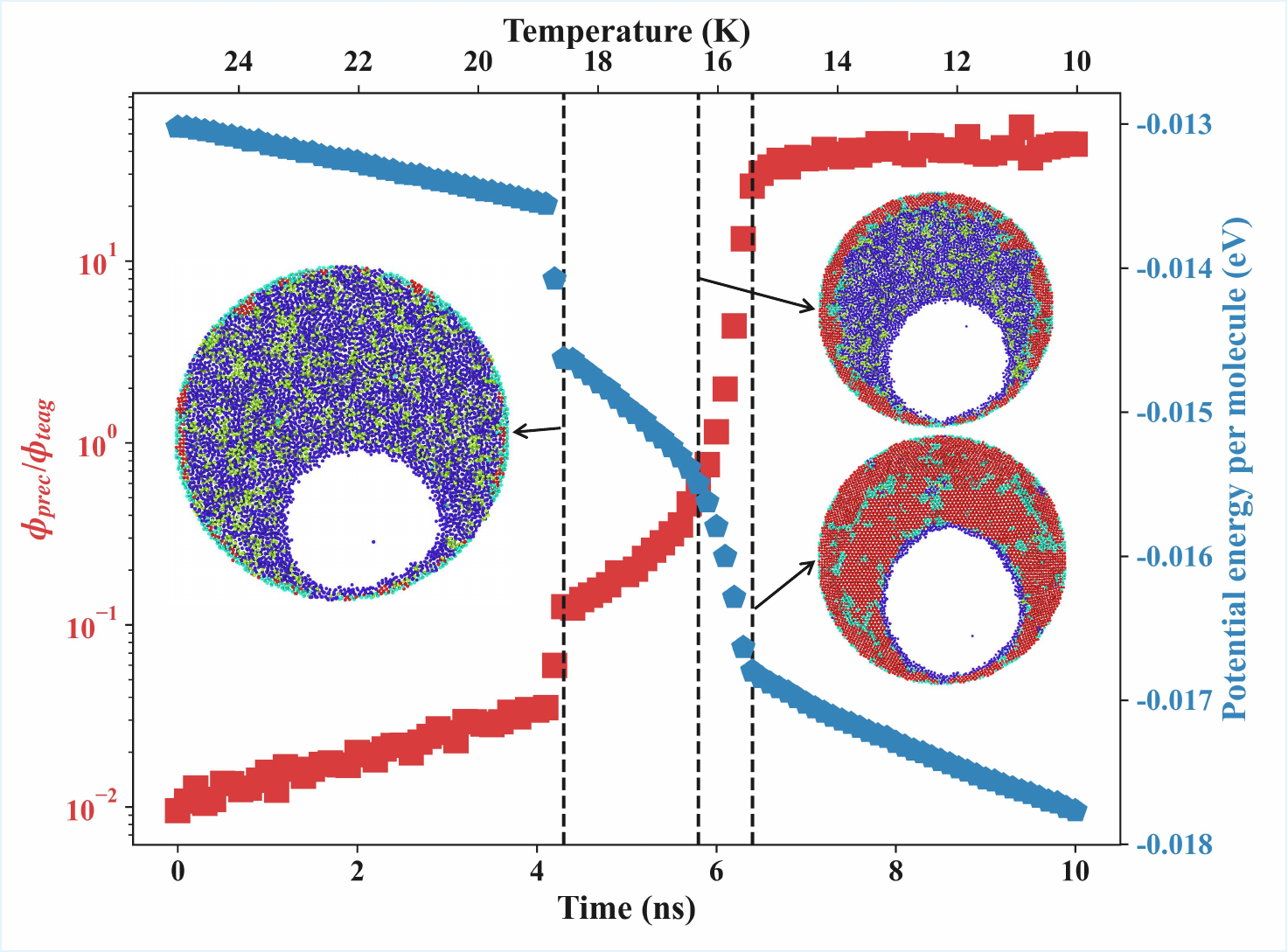} 
	\caption{When the substrate lattice constant is 3.1 Å, this figure shows the ratio of the precursor particle fraction ($\phi_{prec}$) to the geometrically frustrated particle fraction ($\phi_{teag}$) as a function of time. For better comparison, the curve of single-molecule potential energy versus time is also included (displayed as the blue curve). In the crystal growth schematic, red represents solid D$_2$, cyan on the solid surface denotes precursor particles, green indicates geometrically frustrated particles, and blue stands for other liquid D$_2$.} 
	\textbf{Alt text:} Ratio of precursor to frustrated particle fractions over time, together with potential energy curve, for D$_2$ growth on a 3.1 Å substrate. The fastest growth rate of this ratio corresponds to the steepest decrease in potential energy.
	\label{31thermal}
\end{figure}

\begin{figure}
	\centering
	\includegraphics[width=0.52\textwidth]{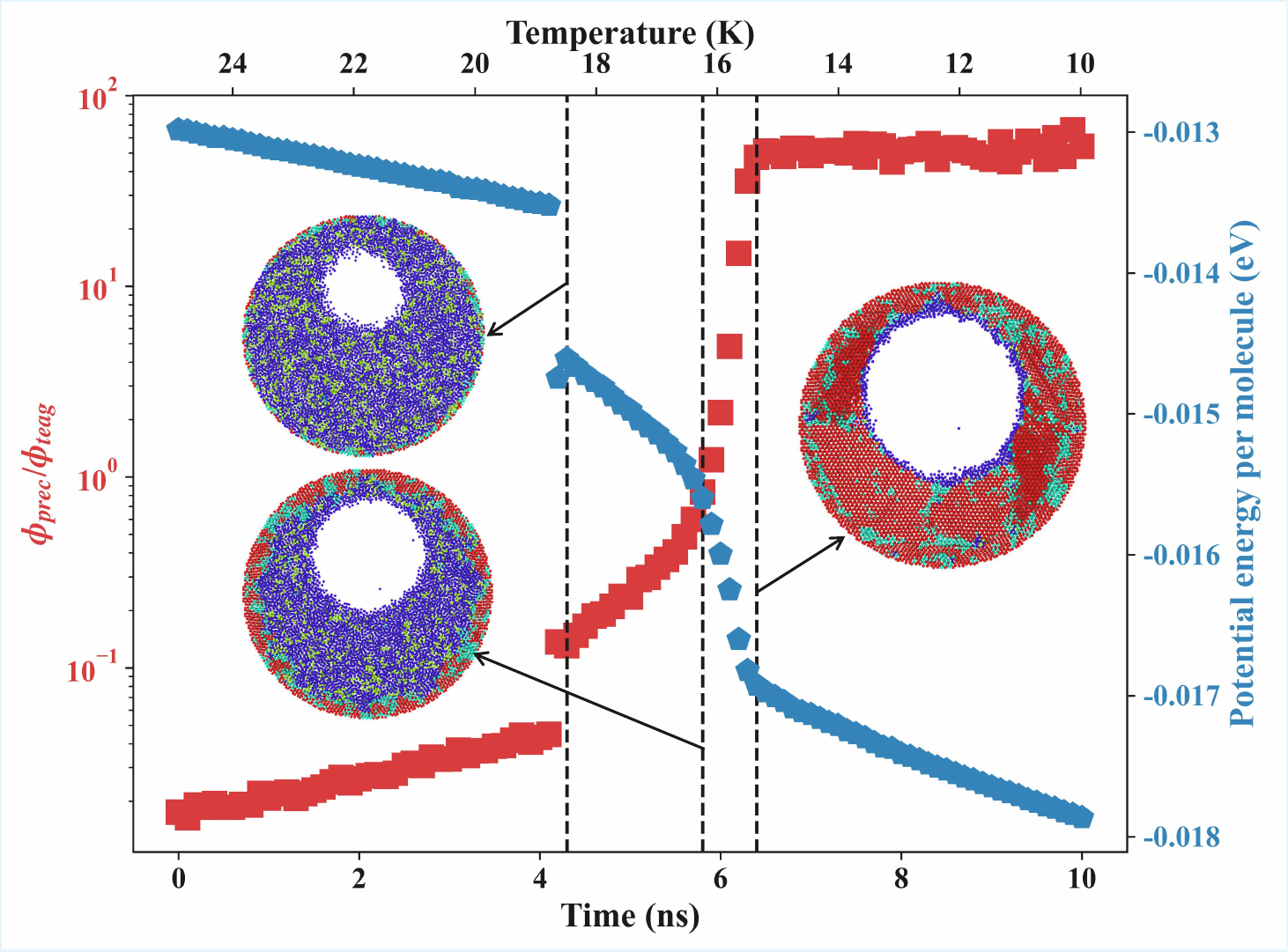} 
	\caption{When the substrate lattice constant is 3.3 Å, this figure shows the ratio of the precursor particle fraction ($\phi_{prec}$) to the geometrically frustrated particle fraction ($\phi_{teag}$) as a function of time. For better comparison, the curve of single-molecule potential energy versus time is also included (displayed as the blue curve). In the crystal growth schematic, red represents solid D$_2$, cyan on the solid surface denotes precursor particles, green indicates geometrically frustrated particles, and blue stands for other liquid D$_2$.} 
	\textbf{Alt text:} Ratio of precursor to frustrated particle fractions over time, together with potential energy curve, for D$_2$ growth on a 3.3 Å substrate. The fastest growth rate of this ratio corresponds to the steepest decrease in potential energy.
	\label{33thermal}
\end{figure}

\begin{figure}
	\centering
	\includegraphics[width=0.52\textwidth]{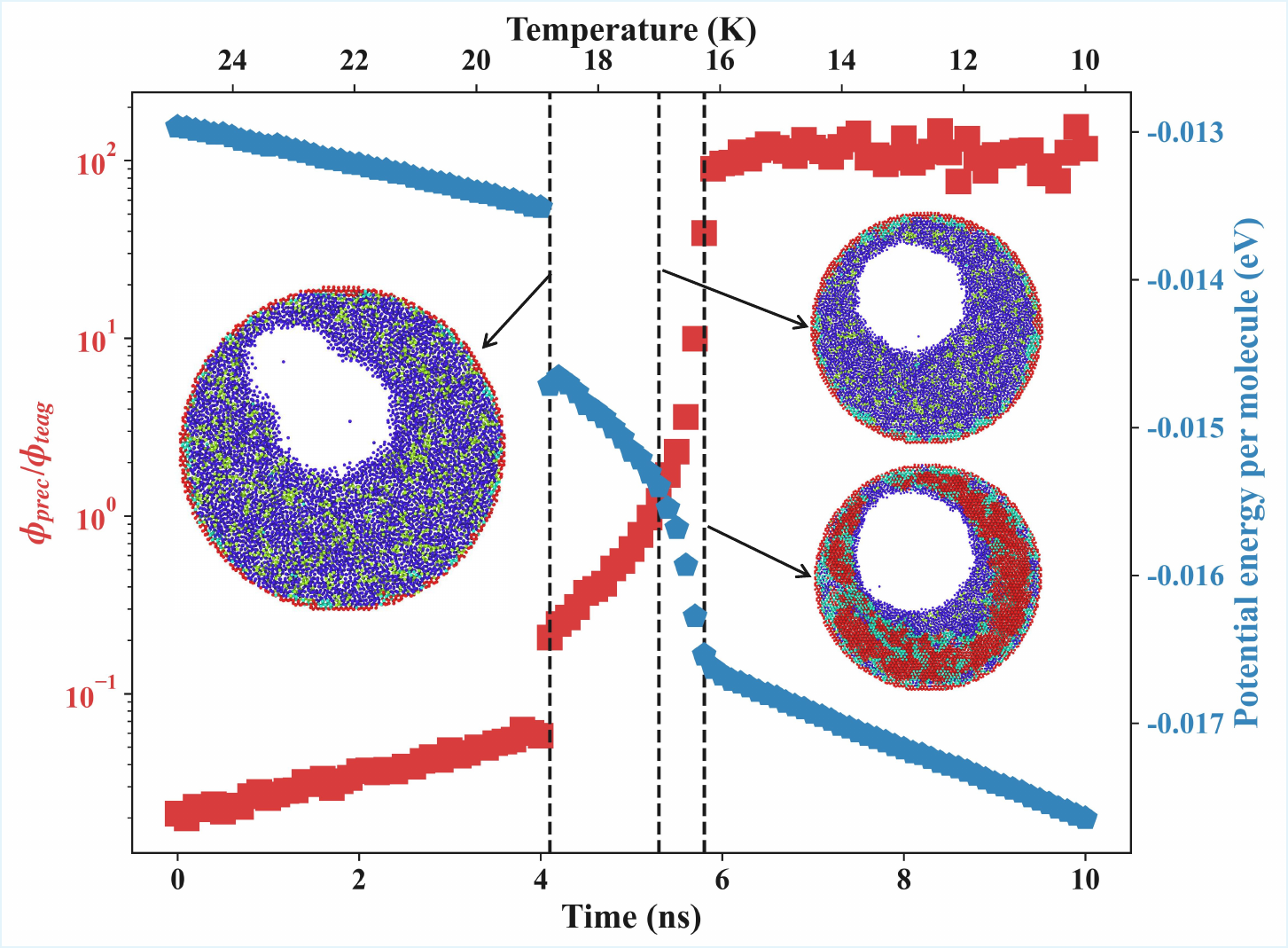} 
	\caption{When the substrate lattice constant is 3.7 Å, this figure shows the ratio of the precursor particle fraction ($\phi_{prec}$) to the geometrically frustrated particle fraction ($\phi_{teag}$) as a function of time. For better comparison, the curve of single-molecule potential energy versus time is also included (displayed as the blue curve). In the crystal growth schematic, red represents solid D$_2$, cyan on the solid surface denotes precursor particles, green indicates geometrically frustrated particles, and blue stands for other liquid D$_2$.} 
	\textbf{Alt text:} Ratio of precursor to frustrated particle fractions over time, together with potential energy curve, for D$_2$ growth on a 3.7 Å substrate. The fastest growth rate of this ratio corresponds to the steepest decrease in potential energy.
	\label{37thermal}
\end{figure}

\begin{figure}
	\centering
	\includegraphics[width=0.52\textwidth]{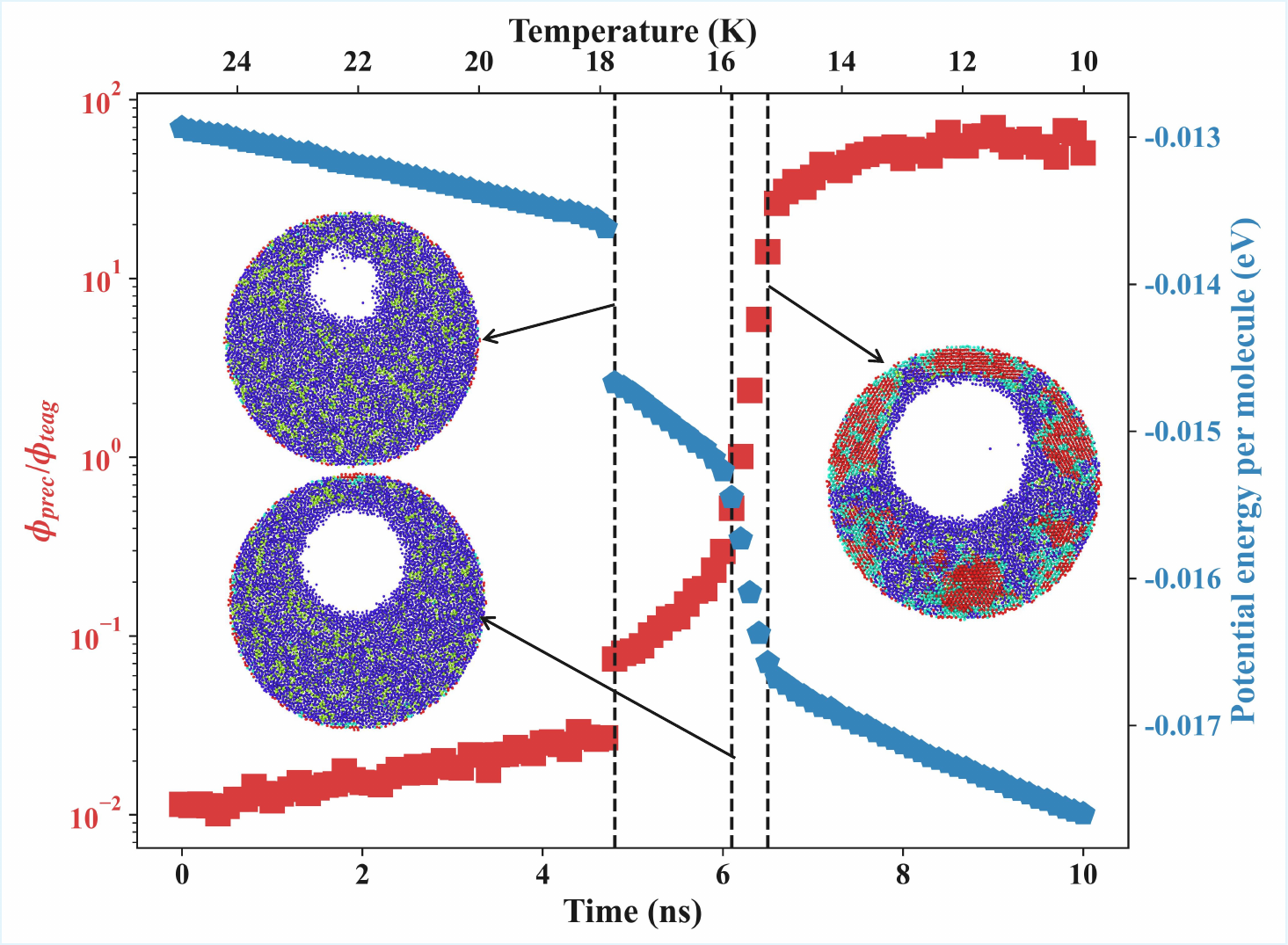} 
	\caption{When the substrate lattice constant is 3.9 Å, this figure shows the ratio of the precursor particle fraction ($\phi_{prec}$) to the geometrically frustrated particle fraction ($\phi_{teag}$) as a function of time. For better comparison, the curve of single-molecule potential energy versus time is also included (displayed as the blue curve). In the crystal growth schematic, red represents solid D$_2$, cyan on the solid surface denotes precursor particles, green indicates geometrically frustrated particles, and blue stands for other liquid D$_2$.} 
	\textbf{Alt text:} Ratio of precursor to frustrated particle fractions over time, together with potential energy curve, for D$_2$ growth on a 3.9 Å substrate. The fastest growth rate of this ratio corresponds to the steepest decrease in potential energy.
	\label{39thermal}
\end{figure}

Fig.~\ref{31thermal}, Fig.~\ref{33thermal}, Fig.~\ref{37thermal} and Fig.~\ref{39thermal} respectively present the time evolution of the ratio between the precursor particle fraction and the geometrically frustrated particle fraction during D$_2$ growth on substrates with lattice constants of 3.1, 3.3, 3.7, and 3.9 Å, all of which exhibit the consistent three-stage evolution characteristic as described in the main text.

\section{Animated video illustrations of the trajectories}
In the compressed package containing the trajectory animations, the prefixes correspond to the substrates with the respective lattice constants, and the suffixes are structure and class. The structure files refer to the classification of particles into HCP, FCC, others and other such categories, while class denotes the classification into precursor particles and geometrically frustrated particles.

\section{Dislocation Density of D$_2$}
\begin{figure}
	\centering
	\includegraphics[width=0.65\textwidth]{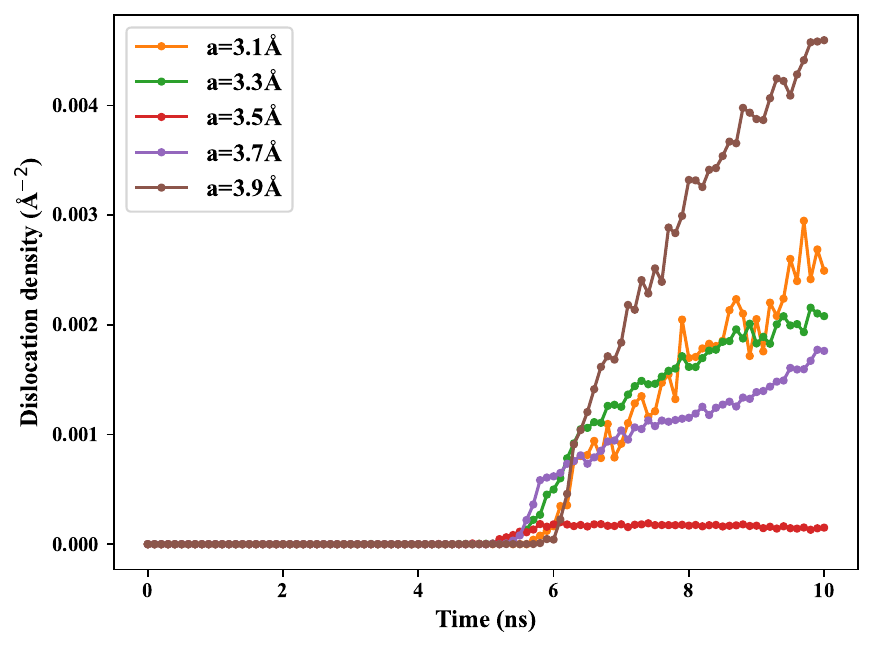} 
	\caption{Dislocation density of D$_2$ crystals grown on substrates with different lattice constants as a function of time.} 
	\textbf{Alt text:} Dislocation density of D$_2$ crystals grown on substrates with different lattice constants versus time. The dislocation density increases significantly with increasing lattice mismatch.
	\label{dxa}
\end{figure}

We analyzed the dislocation density of D$_2$ grown on different substrates after complete crystallization using the Dislocation Analysis (DXA) method implemented in the OVITO software \cite{stukowski2009visualization,stukowski2010extracting,stukowski2012automated}, with the results presented in Fig.~\ref{dxa}. It can be clearly seen that the dislocation density of D$_2$ grown on the substrate with a lattice constant of 3.5 Å is the lowest, and overall, the dislocation density increases gradually with the increase in lattice mismatch.

\section{Diffraction Pattern of D$_2$ After Complete Crystallization}
\begin{figure}
	\centering
	\includegraphics[width=0.9\textwidth]{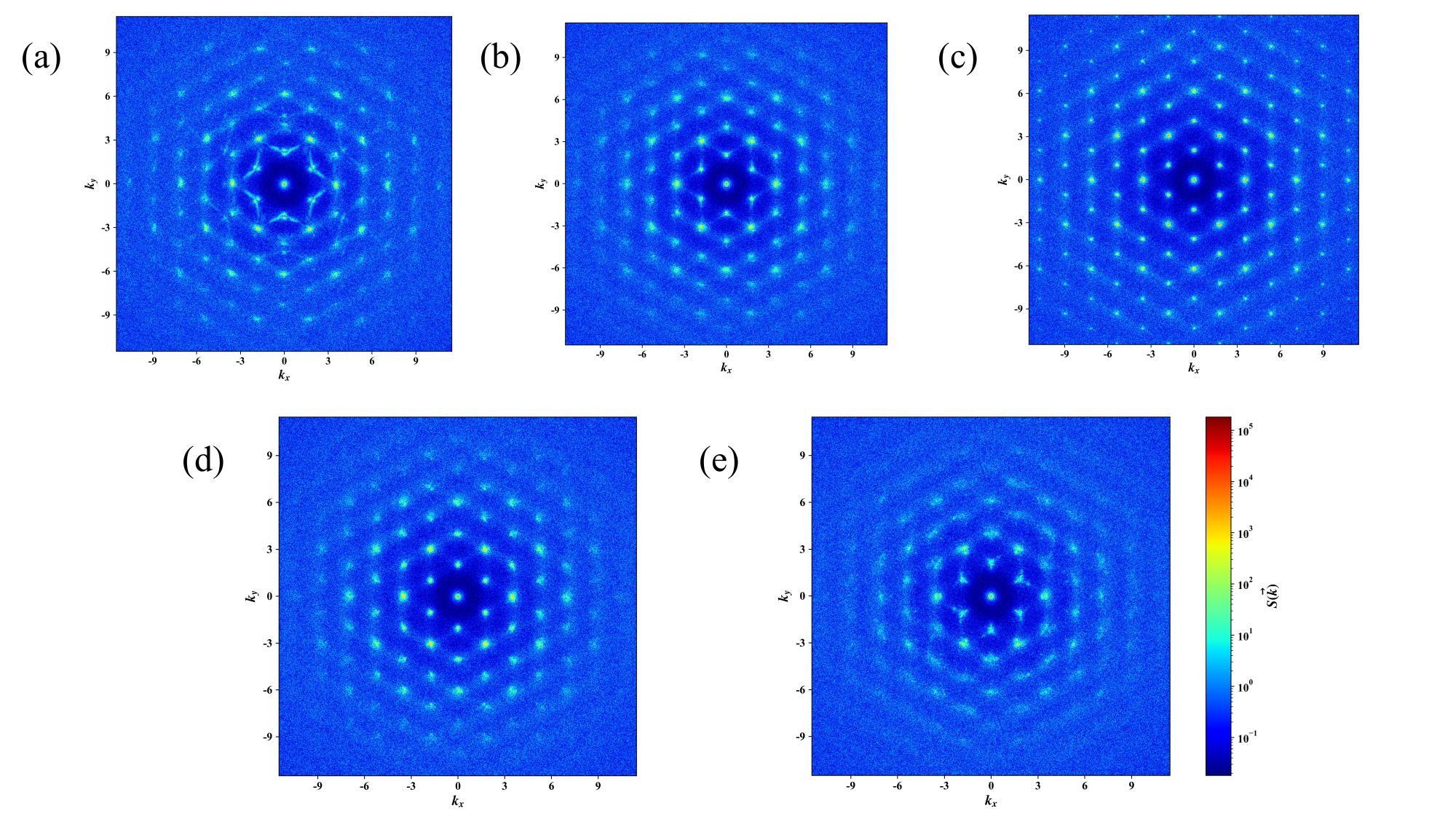} 
	\caption{Diffraction patterns of D$_2$ crystals grown on substrates with different lattice constants: (a) 3.1 Å, (b) 3.3 Å, (c) 3.5 Å, (d) 3.7 Å, and (e) 3.9 Å.} 
	\textbf{Alt text:} Diffraction patterns of D$_2$ crystals grown on substrates with lattice constants from 3.1 Å to 3.9 Å, shown in panels (a)–(e). Only the D$_2$ diffraction pattern grown on the 3.5 Å substrate shows complete HCP symmetry.
	\label{diff}
\end{figure}

The diffraction patterns were calculated using the freud software \cite{ramasubramani2020freud}, which correspond to the 2D images of the static structure factor $S(\vec{k})$ for the atomic point sets. The results are shown in Fig.~\ref{diff}.

\clearpage
\bibliographystyle{apsrev4-1}{}
\bibliography{supp_ref}